\documentclass[a4paper,12pt] {article}
\usepackage{graphics}
\usepackage{graphicx}
\usepackage{mathrsfs}
\usepackage{amssymb}
\usepackage{amsmath}
\usepackage{pstricks}
\usepackage{lmodern}
\usepackage{tikz}
\usetikzlibrary{calc,patterns, decorations.pathmorphing, decorations.markings}
\usepackage{xcolor,rotating,epic,eepic}
\usepackage{pstricks,pst-plot,epstopdf}

\usepackage[latin1]{inputenc}
\usepackage[T1]{fontenc}

\usepackage{color}
\usepackage{pst-all}

\begin{document}

\begin{center}
\Large{\textbf{Nonlinear Generalization of Den Hartog's Equal-Peak Method}}\vspace{1cm}
\end{center}

\begin{center}
{Authors' postprint version\\
Published in : Mechanical systems and signal processing (2014)\\
doi:10.1016/j.ymssp.2014.08.009\\
http://www.sciencedirect.com/science/article/pii/S0888327014003252\\ \vspace{0.2cm}
G. Habib, T. Detroux, R. Vigui\'e, G. Kerschen\\\vspace{0.8cm}

\small Space Structures and Systems Laboratory\\
Department of Aerospace and Mechanical Engineering\\
University of Liege, Liege, Belgium \\
E-mail: giuseppe.habib,tdetroux,g.kerschen@ulg.ac.be\\\vspace{0.5cm} \vspace{1cm}

\rule{0.85\linewidth}{.3pt}
\vspace{-0.5cm}
\begin{abstract} 
This study addresses the mitigation of a nonlinear resonance of a mechanical system. In view of the narrow bandwidth of the classical linear tuned vibration absorber, a nonlinear absorber, termed the nonlinear tuned vibration absorber (NLTVA), is introduced in this paper. An  unconventional aspect of the NLTVA is that the mathematical form of its restoring force is tailored according to the nonlinear restoring force of the primary system. The NLTVA parameters are then determined using a nonlinear generalization of Den Hartog's equal-peak method. The mitigation of the resonant vibrations of a Duffing oscillator is considered to illustrate the proposed developments.

\vspace{1cm}

\end{abstract}
\vspace{-0.5cm}
\rule{0.85\linewidth}{.3pt}

\vspace{0.5cm} Corresponding author: \\ Giuseppe Habib\\
Space Structures and Systems Laboratory\\
Department of Aerospace and Mechanical Engineering\\
University of Liege
\\ 1 Chemin des Chevreuils (B52/3), B-4000 Li\`ege, Belgium. \\
Email: giuseppe.habib@ulg.ac.be.
\vspace{2cm}\\}
\end{center}

\normalsize

\newpage

\section{Introduction}

With continual interest in expanding the performance envelope of engineering systems, nonlinear components are increasingly utilized 
in real-world applications. Mitigating the resonant vibrations of nonlinear structures is therefore becoming a problem of great practical significance \cite{Ibrahim}; 
it is the focus of the present study. It also represents an important challenge, because nonlinear systems exhibit rich and complex phenomena, 
which linear systems cannot. Specifically, one key characteristic of nonlinear oscillations is that their frequency depends intrinsically on motion amplitude.

Nonlinear vibration absorbers, including the autoparametric vibration absorber \cite{Nayfeh,Bajaj}, the nonlinear energy sink (NES) 
\cite{Gend,Book,Lamarque,Cochelin,Gourc} and other variants \cite{Shaw4,Ema,Collette,Poovarodom,Alexander,Lacarbonara}, 
can absorb disturbances in wide ranges of frequencies due to their increased bandwidth. For instance, it was shown that an NES, i.e., 
an essentially nonlinear absorber, can extract energy from virtually any mode of a host structure \cite{NLD}. 
The NES can also carry out targeted energy transfer, which is an irreversible channeling of vibrational energy from the host structure to the absorber \cite{SIAM}. 
This makes nonlinear vibration absorbers suitable candidates for vibration mitigation of nonlinear primary structures. However, the performance of existing nonlinear vibration absorbers is known to exhibit marked sensitivity to motion amplitudes. For instance, 
there exists a well-defined threshold of input energy below which no significant energy dissipation can be induced in an NES \cite{Book}. 
Likewise, the saturation phenomenon --- characteristic of autoparametric vibration absorbers --- occurs only when the forcing amplitude exceeds 
a certain threshold \cite{Nayfeh}. 

This paper builds upon previous developments \cite{Viguie1,Viguie2} to introduce a nonlinear vibration absorber for 
mitigating the vibrations around a nonlinear resonance. The absorber is termed the nonlinear tuned vibration absorber (NLTVA), 
because its nonlinear restoring force is tuned according to the nonlinear restoring force of the host structure. In other words, we propose to synthesize
the absorber's load-deflection characteristic so that the NLTVA can mitigate the considered nonlinear resonance in wide ranges of motion amplitudes. 
In view of the existing literature on nonlinear vibration absorbers, this synthesis represents an unconventional aspect of this work. Interestingly, 
this objective is similar in essence to what was achieved with centrifugal pendulum vibration absorbers in rotating machinery 
\cite{Denman,Shaw2,Shaw3}. Because pendulum absorbers have a natural frequency that scales with the rate of rotation, they can be tuned over 
a continuous range of rotor speeds, e.g., to follow an engine order line. In the same way, Lacarbonara et al. \cite{Lacarbonara2} proposed a 
carefully-tuned secondary pendulating mass in order to reduce the vibrations of a planar pendulum in a relatively large interval of disturbance 
amplitudes. Other studies looked at the influence of absorber nonlinearity on vibration suppression performance. For instance, two nonlinear 
damping mechanisms for a tuned mass damper were compared for the suppression of self-excited oscillations in \cite{Gattulli}. Febbo and Machado \cite{Machado} 
showed that a nonlinear absorber with saturable nonlinearity is more effective than a cubic nonlinearity for vibration mitigation of nonlinear primary 
oscillators. Finally, Agnes suggested to use softening (hardening) absorbers in the presence of hardening (softening) primary systems \cite{Agnes}. 
 
Another contribution of this research is to develop a nonlinear generalization of Den Hartog's equal-peak method which is widely used for designing linear vibration absorbers. 
The basic idea of the nonlinear tuning rule is to select the nonlinear coefficient of the absorber that ensures equal peaks in the nonlinear receptance function for an as large as possible 
range of forcing amplitudes. We will show that this is only feasible when the mathematical form of the NLTVA's restoring force is carefully chosen, which justifies 
the proposed synthesis of the absorber's load-deflection curve.

The paper is organized as follows. Section 2 briefly reviews Den Hartog's equal-peak method and revisits the dynamics of the classical linear tuned 
vibration absorber coupled to a Duffing oscillator. Section 3 lays down the foundations of the NLTVA by proposing a tuning rule for the absorber's 
restoring force. The NLTVA parameters are then determined using a nonlinear generalization of Den Hartog's equal-peak method. The performance 
of the absorber is carefully assessed in Section 4 using a Duffing oscillator as host system. The conclusions of the present study are summarized in Section 5.

\section{The linear tuned vibration absorber (LTVA)}

\subsection{LTVA coupled to a linear oscillator: equal-peak method}

\begin{figure}[t]
\setlength{\unitlength}{1cm}
\begin{picture}(8,8.5)(0,0)
\put(3,0.6)
{\includegraphics[width=10truecm]{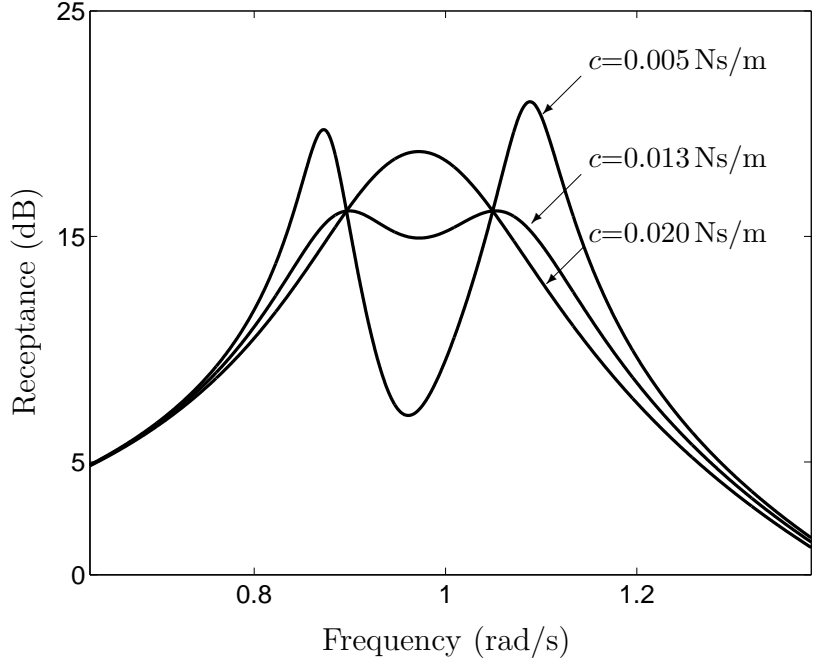}}
\put(6.5,0){Frequency (rad/s)} \put(2.4,3.1){\rotatebox{90}{Receptance (dB)}}
\put(10,7.7){\small{$c$=0.005$\,$Ns/m}}\put(9.9,7.6){\vector(-1,-1){0.5}}
\put(10,6.4){\small{$c$=0.013$\,$Ns/m}}\put(9.9,6.3){\vector(-1,-1){0.65}}
\put(10,5.4){\small{$c$=0.020$\,$Ns/m}}\put(9.95,5.35){\vector(-1,-1){0.5}}
\end{picture}\caption{Illustration of Den Hartog's equal-peak method, $\epsilon=0.05$ and $k_2=0.0454\,$N/m.}\label{DH}
\end{figure}

The steady-state response of an undamped mass-spring system subjected to a harmonic excitation 
at a constant frequency can be suppressed using an undamped linear tuned vibration absorber (LTVA), as proposed by Frahm in 1909 \cite{Frahm}. 
However, the LTVA performance deteriorates significantly when the excitation frequency varies. To improve the performance robustness, damping was 
introduced in the absorber by Ormondroyd and Den Hartog \cite{Ormondroyd}. 
The equations of motion of the coupled system are
\begin{eqnarray}\label{EqLinTVA_0}
   \nonumber
   m_1\ddot{x}_1+k_1x_1+c_2(\dot{x}_1-\dot{x}_2)+k_2(x_1-x_2) & = & F\cos \omega t\\
   m_2\ddot{x}_2+c_2(\dot{x}_2-\dot{x}_1)+k_2(x_2-x_1) & = & 0
\end{eqnarray}
where $x_1(t)$ and $x_2(t)$ are the displacements of the harmonically-forced primary system and of the damped LTVA, respectively. Den Hartog realized that the 
receptance function of the primary mass passes through two invariant points independent of absorber damping. He proposed to adjust the absorber stiffness to have 
two fixed points of equal heights in the receptance curve and to select the absorber damping so that the curve presents a horizontal tangent through one of the fixed points. This laid down the 
foundations of the so-called equal-peak method. Den Hartog \cite{DenHartog} and Brock \cite{Brock} derived approximate analytic formulas for the absorber stiffness and damping, respectively. 
Interestingly, an exact closed-form solution for this classical problem was found only ten years ago \cite{Asami}:
\begin{eqnarray}\label{DHrule}
\nonumber
    \lambda=\frac{\omega_{{n2}}}{\omega_{{n1}}}&=&\sqrt{\frac{k_2m_1}{k_1m_2}}=\frac{2}{1+\epsilon}\sqrt{\frac{2\left[16+23\epsilon+9\epsilon^2+2(2+\epsilon)\sqrt{4+3\epsilon} \right]}
    {3(64+80\epsilon+27\epsilon^2)}}\\
    \mu_2&=&\frac{c_2}{2\sqrt{k_2m_2}}=\frac{1}{4}\sqrt{\frac{8+9\epsilon-4\sqrt{4+3\epsilon}}{1+\epsilon}}
    \end{eqnarray} 
where $\omega_{{n1}}$ and $\omega_{{n2}}$ are the natural frequencies of the primary system and of the absorber, respectively, $\epsilon=m_2/m_1$ is the mass ratio 
and $\mu_2$ is the damping ratio. For $m_1=1\,$kg, $k_1=1\,$N/m and $\epsilon=0.05$, the equal-peak method yields $\lambda=0.952$ and $\mu_2=0.134$, and, 
hence, $k_2=0.0454\,$N/m and $c_2=0.0128\,$Ns/m. As illustrated in Figure \ref{DH}, 
this tuning condition minimizes the maximum response amplitude of the primary system. It is still widely used, as discussed in the review paper \cite{Sun}.

%\begin{figure}
%\begin{centering}
%\includegraphics[trim = 10mm 10mm 10mm 10mm,width=0.32\textwidth]{ClassicalDenHartog_2.eps}
%\par\end{centering}
%\caption{Illustration of Den Hartog's equal-peak method, $\epsilon=0.05$ and $k_2=0.0454\,$N/m.\label{DH}}
%\end{figure}

\subsection{LTVA coupled to a Duffing oscillator}\label{LDO}

The primary system considered throughout this paper is a harmonically-forced, lightly-damped Duffing oscillator. The performance of the LTVA 
attached to this nonlinear system is investigated. The augmented equations of motion are
\begin{eqnarray}\label{EqLinTVA}
   \nonumber
   m_1\ddot{x}_1+c_1\dot{x}_1+k_1x_1+k_{nl1}x_1^3+c_2(\dot{x}_1-\dot{x}_2)+k_2(x_1-x_2) & = & F\cos \omega t\\
   m_2\ddot{x}_2+c_2(\dot{x}_2-\dot{x}_1)+k_2(x_2-x_1) & = & 0
\end{eqnarray}
The parameters are $m_1=1\,$kg, $c_1=0.002\,$Ns/m, $k_1=1\,$N/m and $k_{nl1}=1\,$N/m$^3$. An absorber with a mass ratio $\epsilon$ of 5\% is considered 
for obvious practical reasons. Even if formulas (\ref{DHrule}) are strictly valid only for an undamped primary system, they are still used to determine the 
LTVA parameters in view of the very light damping considered (i.e., damping ratio of 0.1$\,$\%).  

\begin{figure}[p]
\setlength{\unitlength}{1cm}
\begin{picture}(8,13.5)(0,0)
\put(0.75,7){\includegraphics[width=7.2truecm]{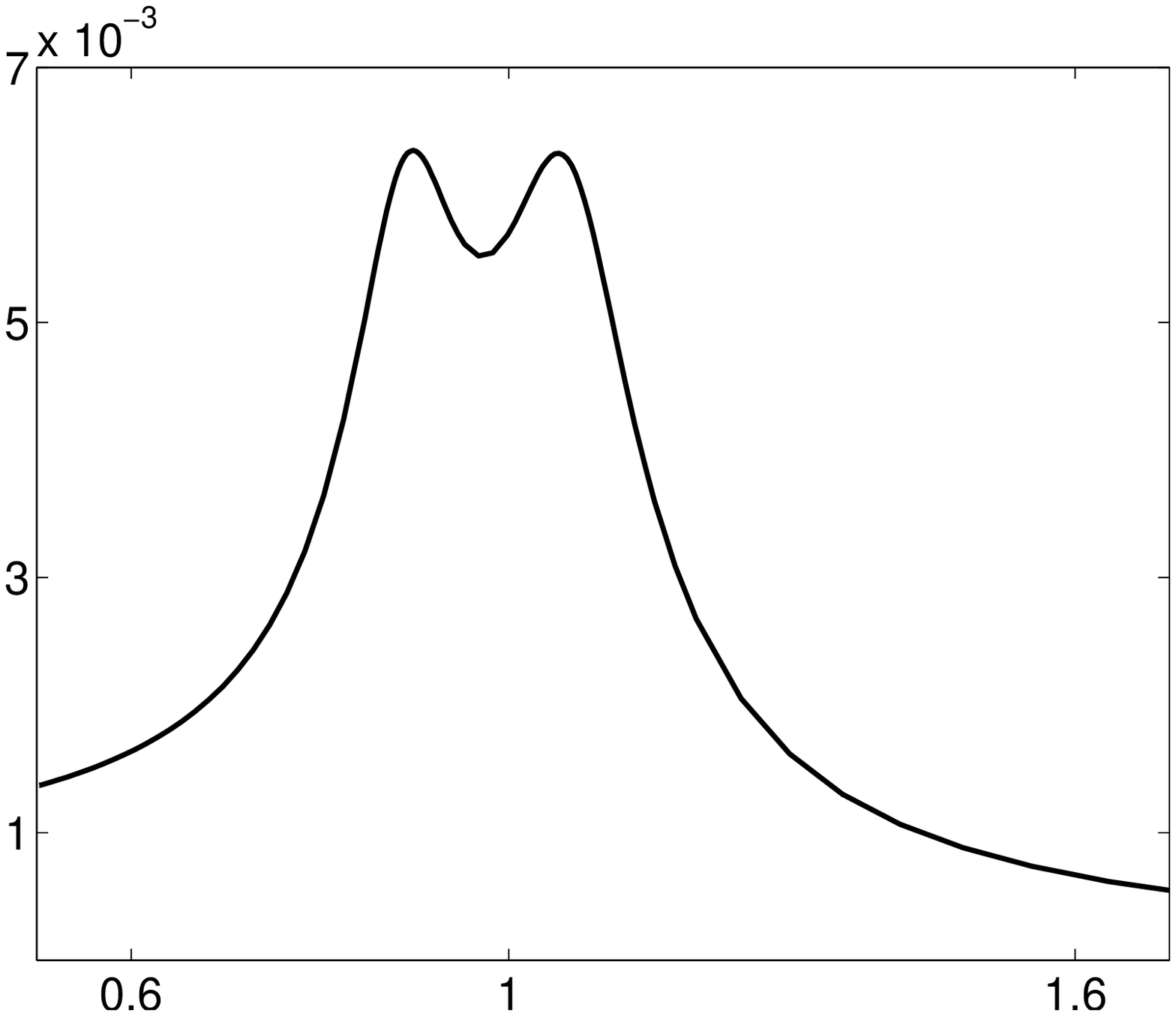}}
\put(8.45,7){\includegraphics[width=7.5truecm,height=5.9truecm]{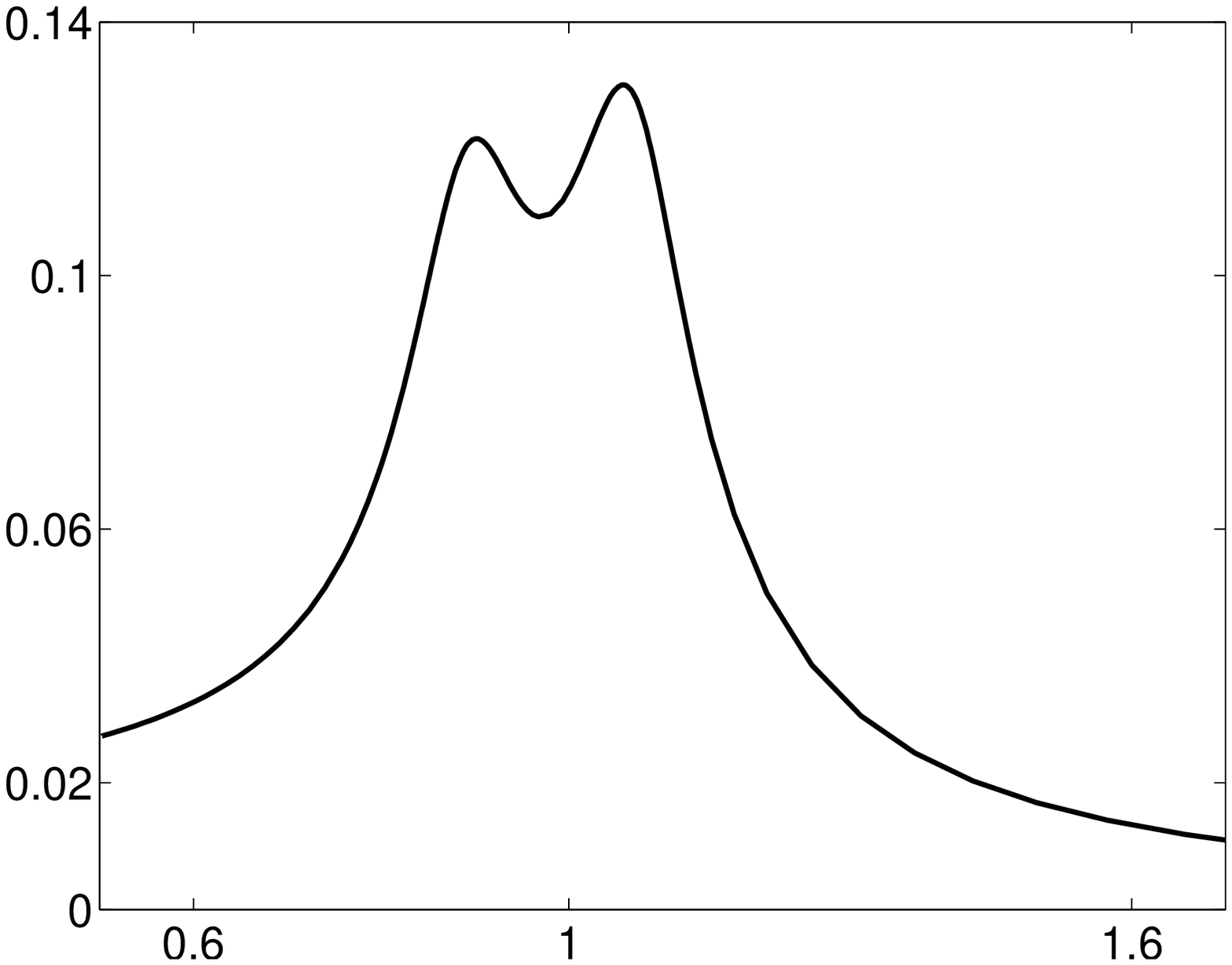}}
\put(0.5,0.6){\includegraphics[width=7.5truecm]{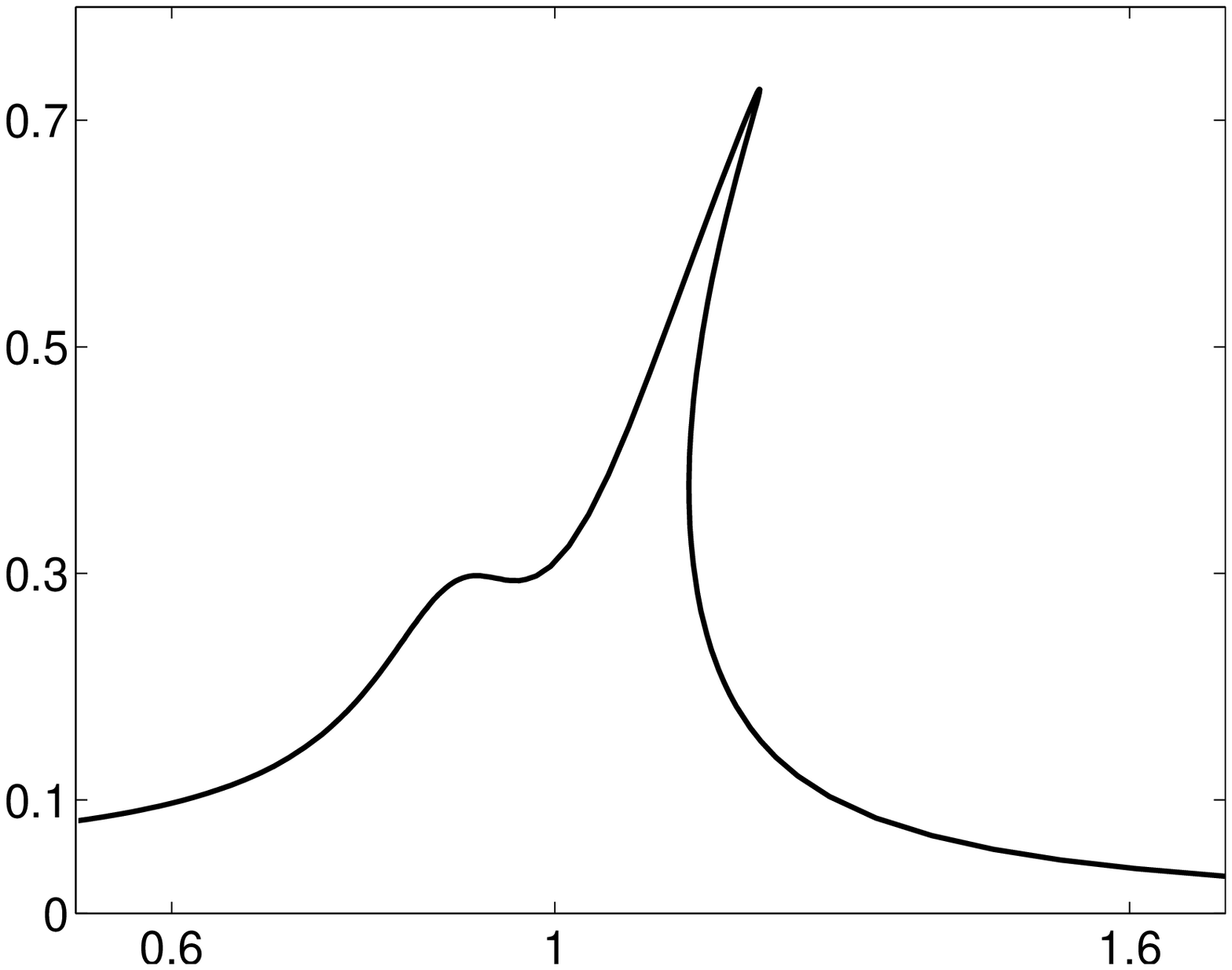}}
\put(8.5,0.6){\includegraphics[width=7.5truecm]{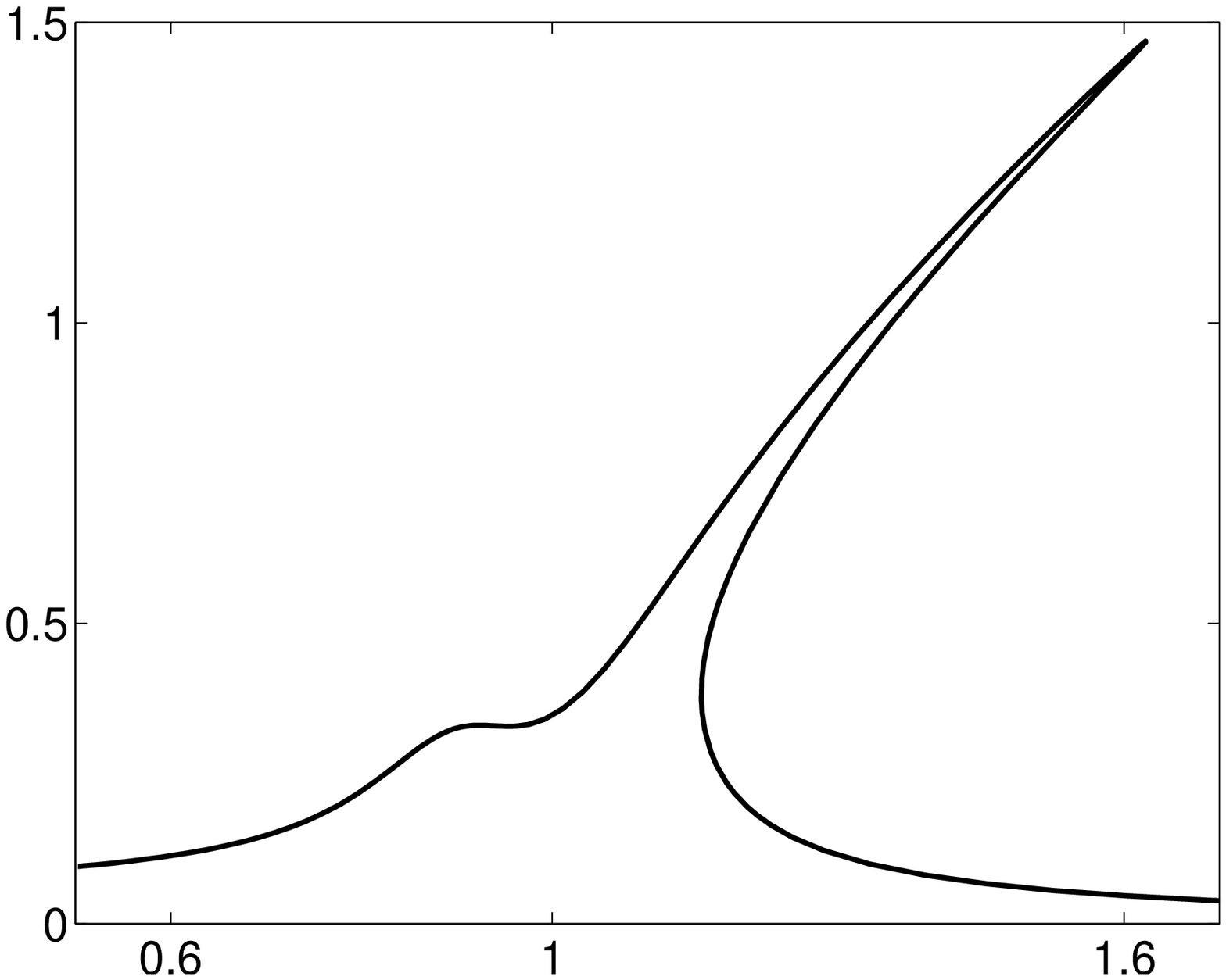}}
\put(6.8,0){Frequency (rad/s)} \put(-0.1,5.3){\rotatebox{90}{Displacement
(m)}}
 \put(1.2,12){(a)}
 \put(9.3,12){(b)}
 \put(1.2,5.7){(c)}
 \put(9.3,5.7){(d)}
\end{picture}\caption{Frequency response of a Duffing oscillator with an attached LTVA. (a) $F=0.001\,$N; (b) $F=0.02\,$N; (c) $F=0.06\,$N, and (d) $F=0.07\,$N.}\label{LINTVA}
\end{figure}

\begin{figure}[p]
\setlength{\unitlength}{1cm}
\begin{picture}(8,6.5)(0,0)
\put(4.5,0.6){\includegraphics[width=7.5truecm]{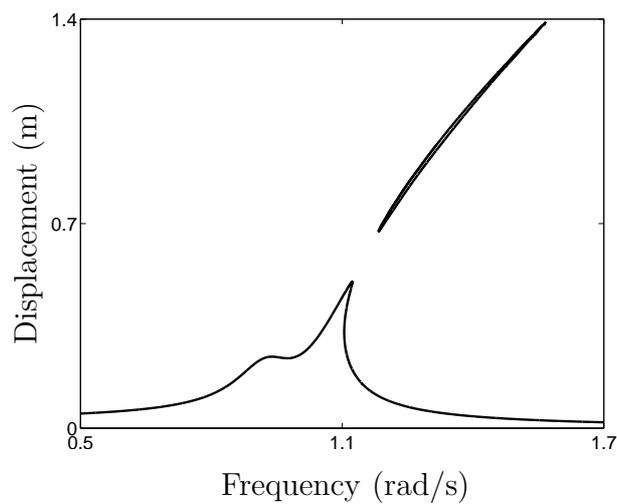}}
\put(6.75,0){Frequency (rad/s)} \put(4,2.){\rotatebox{90}{Displacement
(m)}}
 \end{picture}\caption{Frequency response of a Duffing oscillator with an attached LTVA ($\epsilon=0.03$, $F=0.035\,$N). }\label{LTVA_Isola}
\end{figure}

Figure \ref{LINTVA} shows the displacement response of the primary mass for various forcing amplitudes $F$ ranging from $0.001\,$N to $0.07\,$N. 
Because the dynamics of a nonlinear system is considered, these curves were computed using a path-following algorithm combining shooting and 
pseudo-arclength continuation. The algorithm is similar to that used in \cite{Peeters} for the computation of nonlinear normal modes. 
At 0.001$\,$N in Figure \ref{LINTVA}(a), the system behaves linearly, and two peaks of equal amplitude can be observed in accordance with linear theory. 
When the forcing amplitude is increased, the cubic nonlinearity of the primary system is activated. At 0.02$\,$N, Figure \ref{LINTVA}(b) 
shows a slight detuning of the absorber, but this detuning is not yet too detrimental to absorber performance. At 0.06$\,$N in Figure \ref{LINTVA}(c), 
the LTVA is no longer effective due to the important difference in the amplitude of the two resonances. A hardening behavior characteristic of cubic 
springs with positive coefficients is also present in the second resonance peak; it indicates that the regime of motion is no longer weakly nonlinear. 
When the forcing amplitude increases from $0.06\,$N to $0.07\,$N, the LTVA becomes completely detuned by the nonlinear effects. 
The primary mass displacement increases by a factor of 2 between Figures \ref{LINTVA}(c) and \ref{LINTVA}(d), 
a clear sign of the absence of superposition principle for this coupled system. Clearly, in view of the frequency-energy dependence of 
nonlinear oscillations and of the narrow bandwidth of the LTVA, this absorber can only be effective in weakly nonlinear regimes of motion. Moreover, 
due to the non-uniqueness of nonlinear solutions, additional resonances can be observed when a LTVA is coupled to a nonlinear system. This is depicted 
in Figure \ref{LTVA_Isola} where a detached resonance curve appears for a lower mass ratio of 3\% and a forcing amplitude of $0.035\,$N.

\section{The nonlinear tuned vibration absorber (NLTVA)}

In view of the results presented in the previous section, it is meaningful to examine the performance of 
nonlinear absorbers for vibration mitigation of nonlinear primary structures. Roberson was the first to 
observe a broadening of the suppression band through the addition of a nonlinear spring that he chose to be cubic for 
facilitating its practical realization \cite{Roberson}. However, as pointed out in the introductory section, 
this increased bandwidth may come at the price of a marked sensitivity to external forcing amplitude.

To mitigate a nonlinear resonance in an as large as possible range of forcing amplitudes, we introduce the nonlinear tuned vibration absorber (NLTVA). 
One unconventional feature of this absorber is that the mathematical form of its nonlinear restoring force is not imposed a priori, as it is the case for most 
existing nonlinear absorbers. Instead, we propose to fully exploit the additional design parameter offered by nonlinear devices and, hence, to synthesize the absorber's load-deflection 
curve according to the nonlinear restoring force of the primary structure. 

\subsection{Synthesis of the nonlinear restoring force of the absorber}\label{synth}

\begin{figure}
\begin{centering}
\begin{tikzpicture}[scale=1.2]
\tikzstyle{spring}=[thick,decorate,decoration={zigzag,pre length=0.3cm,post length=0.3cm,segment length=6}]
\tikzstyle{damper}=[thick,decoration={markings,
  mark connection node=dmp,
  mark=at position 0.5 with
  {
    \node (dmp) [thick,inner sep=0pt,transform shape,rotate=-90,minimum width=15pt,minimum height=5pt,draw=none] {};
    \draw [thick] ($(dmp.north east)+(2pt,0)$) -- (dmp.south east) -- (dmp.south west) -- ($(dmp.north west)+(2pt,0)$);
    \draw [thick] ($(dmp.north)+(0,-5pt)$) -- ($(dmp.north)+(0,5pt)$);
  }
}, decorate]
%\tikzstyle{ground}=[pattern=forGround,hatchspread=7pt,hatchthickness=0.7pt, draw=none,minimum width=0.75cm,minimum height=0.3cm]
%
%\draw[ground] (-.5,1)rectangle(0,-1);
\draw[thick] (0,1)--(0,-1);
\draw[spring] (0,0)--(2,0);
\node[above left] at (1,0) {$k_{1}$};
\draw[spring] (0,-.7)--(2,-.7);
\draw[thick,->] (.5,-1.1)--(1.7,-.4);
\node[above left] at (1,-.7) {$k_{nl1}$};
\draw[damper] (0,0.7)--(2,0.7);
\node[above left] at (1,0.7) {$c_1$};
\draw (2,1)rectangle(4,-1);
\node at (3,0) {$m_1$} ;
\draw[->] (3,1)|-(3.5,1.5);
\node[above] at (3.5,1.5) {$x_1$};
%\draw[thick,->] (1.8,1.2)--(2.5,1.2);
%\node[above] at (2,1.2) {$f\cos\omega t$};
%
\draw[damper] (4,0.7)--(6,0.7);
\node[above left] at (5,0.7) {$c_2$};
\draw[-] (4,-.35)--(4.3,-.35);
\draw[-] (5.7,-.35)--(6,-.35);
\draw (4.3,0)rectangle(5.7,-.7);
\node at (5,-0.35) {$?$} ;
\node at (5,-1) {$g(\bullet)$} ;
\draw (6,0.8)rectangle(7,-.8);
\node at (6.5,0) {$m_2$} ;
\draw[->] (6.5,.8)|-(7,1.3);
\node[above] at (7,1.3) {$x_2$};
\end{tikzpicture}
\par\end{centering}
\caption{Schematic representation of an NLTVA attached to a Duffing oscillator.}\label{2dofsystem}
\end{figure}
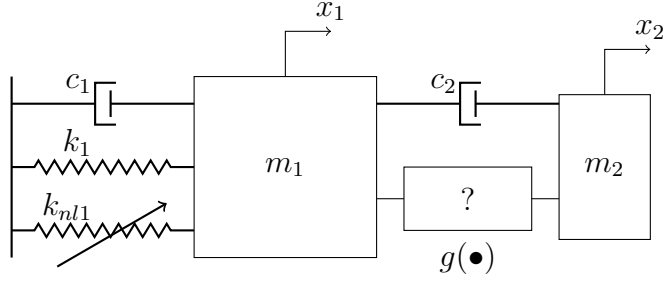

The dynamics of a Duffing oscillator with an attached NLTVA as depicted in Figure \ref{2dofsystem} is considered:
\begin{eqnarray}\label{EqOrig}
   \nonumber
   m_1\ddot{x}_1+c_1\dot{x}_1+k_1x_1+k_{nl1}x_1^3+c_2(\dot{x}_1-\dot{x}_2)+g(x_1-x_2)& = & F\cos \omega t\\
   m_2\ddot{x}_2+c_2(\dot{x}_2-\dot{x}_1)-g(x_1-x_2) & = & 0
\end{eqnarray}

The NLTVA is assumed to have a generic smooth restoring force $g\left(x_1-x_2\right)$ with $g(0)=0$. 
After the definition of the dimensionless time $\tau=\omega_{n1}t$, where $\omega_{n1}=\sqrt{k_1/m_1}$, and the application of the transformation $r(t)=x_1(t)-x_2(t)$ yields
\begin{eqnarray}\label{EqAdim1}
   \nonumber
   x_1''+2\mu_1x_1'+x_1+\frac{4}{3}\tilde\alpha_3x_1^3+2\mu_2\lambda\epsilon r'+\frac{\epsilon}{m_2\omega_{n1}^2}g\left(r\right)=f\cos\gamma\tau\\
   r''+2\mu_1x_1'+x_1+\frac{4}{3}\tilde\alpha_3x_1^3+2\mu_2\lambda \left(\epsilon+1\right)r'+\frac{\epsilon+1}{m_2\omega_{n1}^2}g\left(r\right)=f\cos\gamma\tau
\end{eqnarray}
where prime denotes differentiation with respect to time $\tau$, $2\mu_1=c_1/(m_1\omega_{n1})$, $\tilde\alpha_3 =3/4k_{nl1}/k_1$, $2\mu_2=c_2/(m_2\omega_{n2})$, $\lambda=\omega_{n2}/\omega_{n1}$, $\epsilon=m_2/m_1$, $f=F/k_1$ and $\gamma=\omega/\omega_{n1}$. We note that $\omega_{n2}$ is the linearized frequency of the NLTVA.

Expanding $g(r)$ in Taylor series around $r=0$ and normalizing the system using $q_1=x_1/f$ and $q_2=r/f$, we obtain
\small
\begin{eqnarray}\label{EqAdim2}
\nonumber
q_1''+2\mu_1q_1'+q_1+\frac{4}{3}\tilde\alpha_3f^2q_1^3+2\mu_2\lambda\epsilon q_2'+\lambda^2\epsilon q_2+\frac{\epsilon}{m_2\omega_{n1}^2}\sum_{k=2}^{\infty}\frac{f^{k-1}}{k!}\frac{\text d^kg}{\text dr^k}\bigg|_{r=0}q_2^k=\cos\gamma\tau\\
q_2''+2\mu_1q_1'+q_1+\frac{4}{3}\tilde\alpha_3f^2q_1^3+2\mu_2\lambda\left(\epsilon+1\right)q_2'+\lambda^2\left(\epsilon+1\right)q_2+\frac{\epsilon+1}{m_2\omega_{n1}^2}\sum_{k=2}^{\infty}\frac{f^{k-1}}{k!}\frac{\text d^kg}{\text dr^k}\bigg|_{r=0}q_2^k=\cos\gamma\tau \label{taylor}
\end{eqnarray}
\normalsize
where $\omega_{n2}=\sqrt{\text dg/\text dq_2|_{q_2=0}/m_2}$. 

In Equations (\ref{EqAdim2}), the linear terms are independent of the forcing amplitude $f$, which confirms that a purely linear absorber attached to a linear oscillator 
is effective irrespective of the considered forcing amplitude. Focusing now on the complete system, $f$ appears in the nonlinear coefficients 
of both the primary system and the absorber, which reminds that it is equivalent to consider the system strongly nonlinear or strongly excited. Specifically, Equations (\ref{EqAdim2})
show that the forcing amplitude modifies linearly the quadratic terms, quadratically the cubic terms 
and so on. This suggests that, if an optimal set of absorber parameters is chosen for a specific value of $f$, variations of $f$ will detune the nonlinear absorber, unless the nonlinear 
coefficients of the primary system and of the absorber undergo a similar variation with $f$. According to Equations (\ref{EqAdim2}), this can be achieved by selecting the 
same mathematical function for the absorber as that of the primary system. When coupled to a Duffing oscillator, the NLTVA should therefore possess a cubic spring: 
\begin{eqnarray}\label{EqAdim3}
\nonumber
q_1''+2\mu_1q_1'+q_1+\frac{4}{3}\alpha_3q_1^3+2\mu_2\lambda\epsilon q_2'+\lambda^2\epsilon q_2+\frac{4}{3}\epsilon\beta_3 q_2^3=\cos\gamma\tau\\
q_2''+2\mu_1q_1'+q_1+\frac{4}{3}\alpha_3q_1^3+2\mu_2\lambda\left(\epsilon+1\right)q_2'+\lambda^2\left(\epsilon+1\right)q_2+\frac{4}{3}\left(\epsilon+1\right)\beta_3q_2^3=\cos\gamma\tau \label{taylor2}
\end{eqnarray}
where 
\begin{equation}
  \alpha_3=\tilde\alpha_3f^2 \quad \mbox{and}\quad \beta_3=\frac{3}{4}\frac{f^2g'''(r)|_{r=0}}{3!m_2\omega_{n1}^2}.
\end{equation}
The NLTVA spring should also possess a linear component so that it is effective at low forcing amplitudes 
where the cubic component of the Duffing oscillator is not activated.

In summary, the proposed nonlinear tuning rule is {\it to choose the mathematical form of the NLTVA's restoring force so that it is a `mirror' of 
the primary system.}

\subsection{Nonlinear generalization of the equal-peak method}

The next objective is to determine the NLTVA parameters, namely $\epsilon$, $\lambda$, $\mu_2$ and $\beta_3$. The mass ratio $\epsilon$ is chosen 
according to practical constraints. The linear parameters $\lambda$ and $\mu_2$ are determined using Equations (\ref{DHrule}). 
Because an exact analytic estimation of $\beta_3$ is not within reach, an approximate solution is sought using the harmonic balance method limited 
to one harmonic component, $q_1=A_1\cos\gamma\tau+B_1\sin\gamma\tau$ and $q_2=A_2\cos\gamma\tau+B_2\sin\gamma\tau$. 
Substituting this {\it ansatz} in Equations (\ref{EqAdim3}), applying the approximations $\cos^3\gamma\tau\approx3/4\cos\gamma\tau$ and 
$\sin^3\gamma\tau\approx3/4\sin\gamma\tau$, and balancing cosine and sine terms, the system of polynomial equations 
\small
\begin{eqnarray}
&&\alpha_3A_1^3+\alpha_3A_1B_1^2+\beta_3\epsilon A_2^3+\beta_3\epsilon A_2B_2^2+\left(1-\gamma^2\right)A_1+2\mu_1\gamma B_1+\lambda^2\epsilon A_2+2\mu_2\lambda\epsilon\gamma B_2=1 \nonumber\\
&&\alpha_3A_1^3+\alpha_3A_1B_1^2+\beta_3\left(\epsilon+1\right) A_2^3+\beta_3\left(\epsilon+1\right) A_2B_2^2+A_1+2\mu_1\gamma B_1+\left(\lambda^2\left(\epsilon+1\right)-\gamma^2\right) A_2+\nonumber\\
&&\quad 2\mu_2\lambda\left(\epsilon+1\right)\gamma B_2=1\nonumber\\
&&\alpha_3A_1^2B_1+\alpha_3B_1^3+\beta_3\epsilon A_2^2B_2+\beta_3\epsilon B_2^3-2\mu_1\gamma A_1+\left(1-\gamma^2\right)B_1-2\mu_2\lambda\epsilon\gamma A_2+ \lambda^2\epsilon B_2=0\nonumber\\
&&\alpha_3A_1^2B_1+\alpha_3B_1^3+\beta_3\left(\epsilon+1\right) A_2^2B_2+\beta_3\left(\epsilon+1\right) B_2^3-2\mu_1\gamma A_1+B_1 -2\mu_2\lambda\left(\epsilon+1\right)\gamma A_2 +\nonumber\\
&&\quad\left(\lambda^2\left(\epsilon+1\right)-\gamma^2\right) B_2 =0 \label{HB}
\end{eqnarray}
\normalsize
is obtained. This system is solved for fixed values of $\mu_1=0.001$, $\mu_2=0.134$, $\lambda=0.952$, for different values of $\epsilon$ and $\alpha_3$, and 
for a range of excitation frequencies $\gamma$ encompassing the system's resonances. Starting with weakly nonlinear regimes, i.e., $\alpha_3>0$, 
we seek the value of $\beta_3$, which gives two resonance peaks of equal amplitude. The procedure is repeated for increasing values of $\alpha_3$, which allows 
to consider stronger and stronger nonlinear regimes of motion. 

\begin{figure}[t]
\setlength{\unitlength}{1cm}
\begin{picture}(8,6.4)(0,0)
\put(0.5,0.5){\includegraphics[width=7.5truecm]{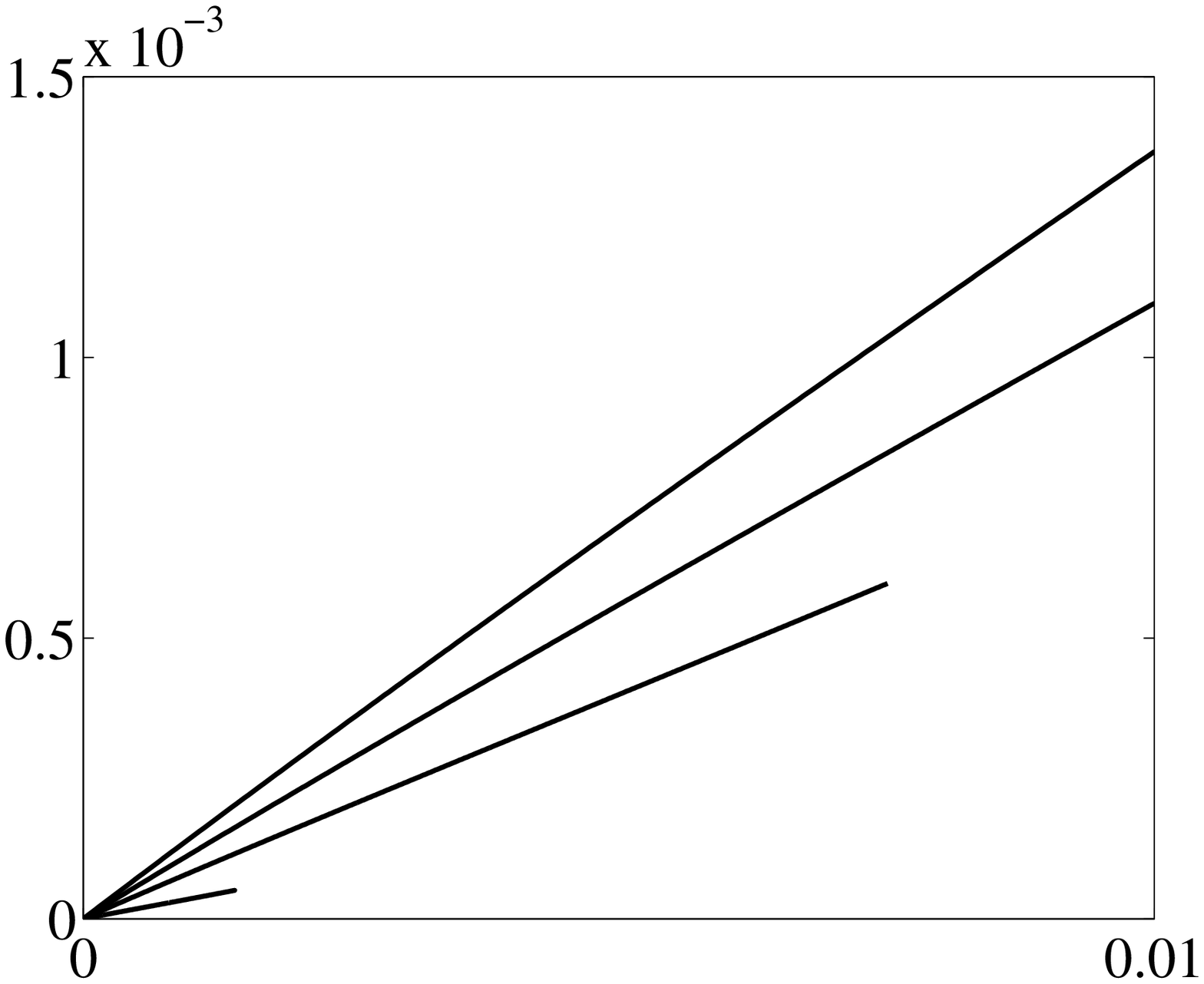}}
\put(8.5,0.5){\includegraphics[width=7.3truecm]{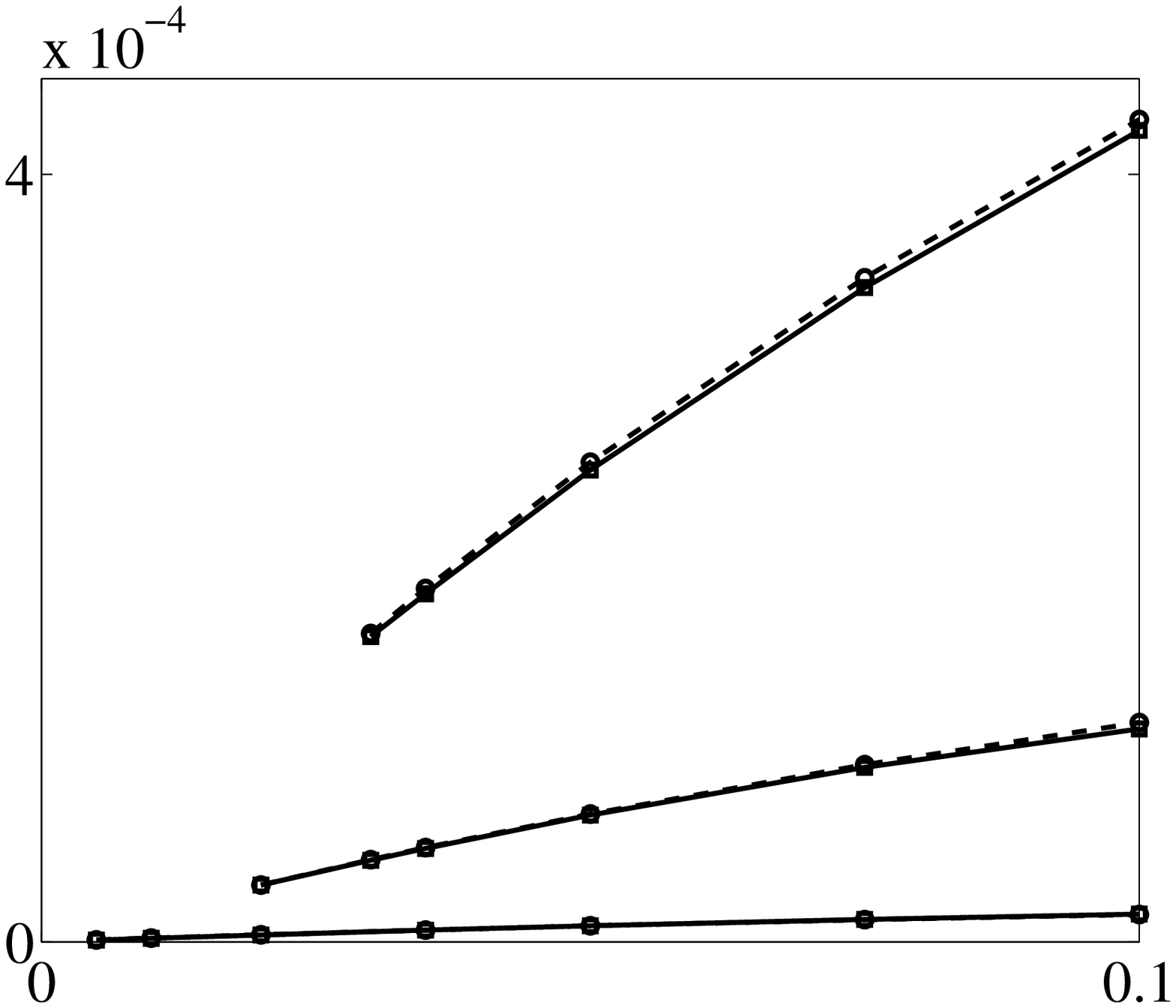}}
\put(4.1,0.5){$\alpha_3$} \put(0.4,3.3){\rotatebox{90}{$\beta_3$}}
\put(12.1,0.5){$\epsilon$} \put(8.2,3.3){\rotatebox{90}{$\beta_3$}}
  \put(1.2,5.7){(a)}
 \put(8.9,5.7){(b)}
 \small
 \put(5.8,5.3){$\epsilon=0.1$}
 \put(5.8,3.5){$\epsilon=0.075$}
  \put(5.8,2.65){$\epsilon=0.05$}
  \put(2.1,1.1){$\epsilon=0.02$}
  \put(13.4,5.85){$\alpha_3=0.003$}
  \put(13.4,2.25){$\alpha_3=8.10^{-4}$}
  \put(13.4,1.25){$\alpha_3=10^{-5}$}
 %\put(1.45,1.1){\line(-1,1){0.5}}
 \normalsize
\end{picture}\caption{Nonlinear equal-peak method. (a) Values of $\beta_3$ realizing equal peaks for increasing $\alpha_3$ and 
different $\epsilon$; (b) values of $\beta_3$ realizing equal peaks for increasing $\epsilon$ and different $\alpha_3$; the solid line 
is the result of the numerical computations, and the dashed line is the regression $\beta_3=2\alpha_3\epsilon/(1+4\epsilon)$.}\label{NL_DH}
\end{figure}

\begin{figure}[t]
\setlength{\unitlength}{1cm}
\begin{picture}(8,6.2)(0,0.3)
\put(0.4,0.6){\includegraphics[width=7.3truecm]{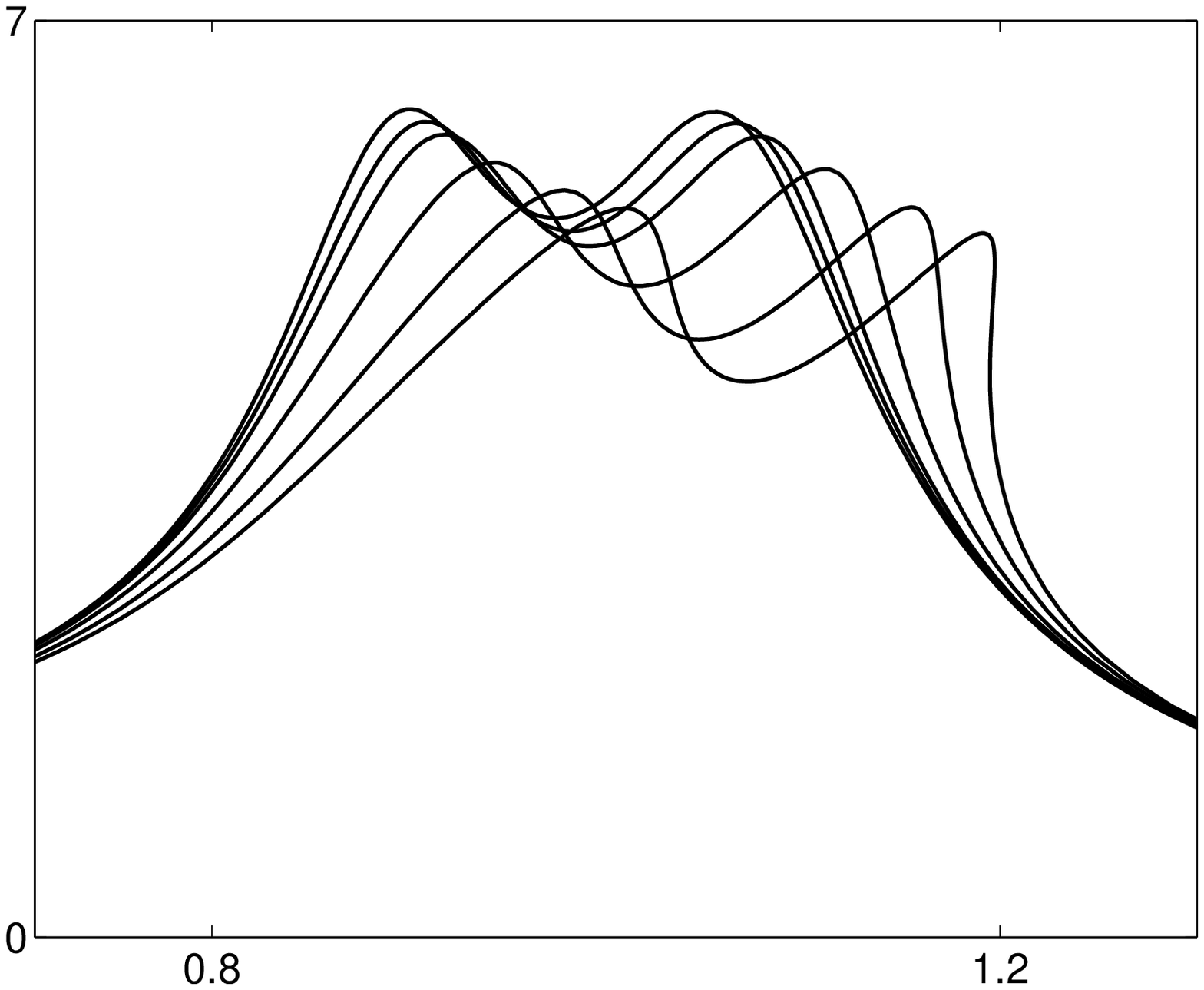}}
\put(8.7,0.6){\includegraphics[width=7.3truecm]{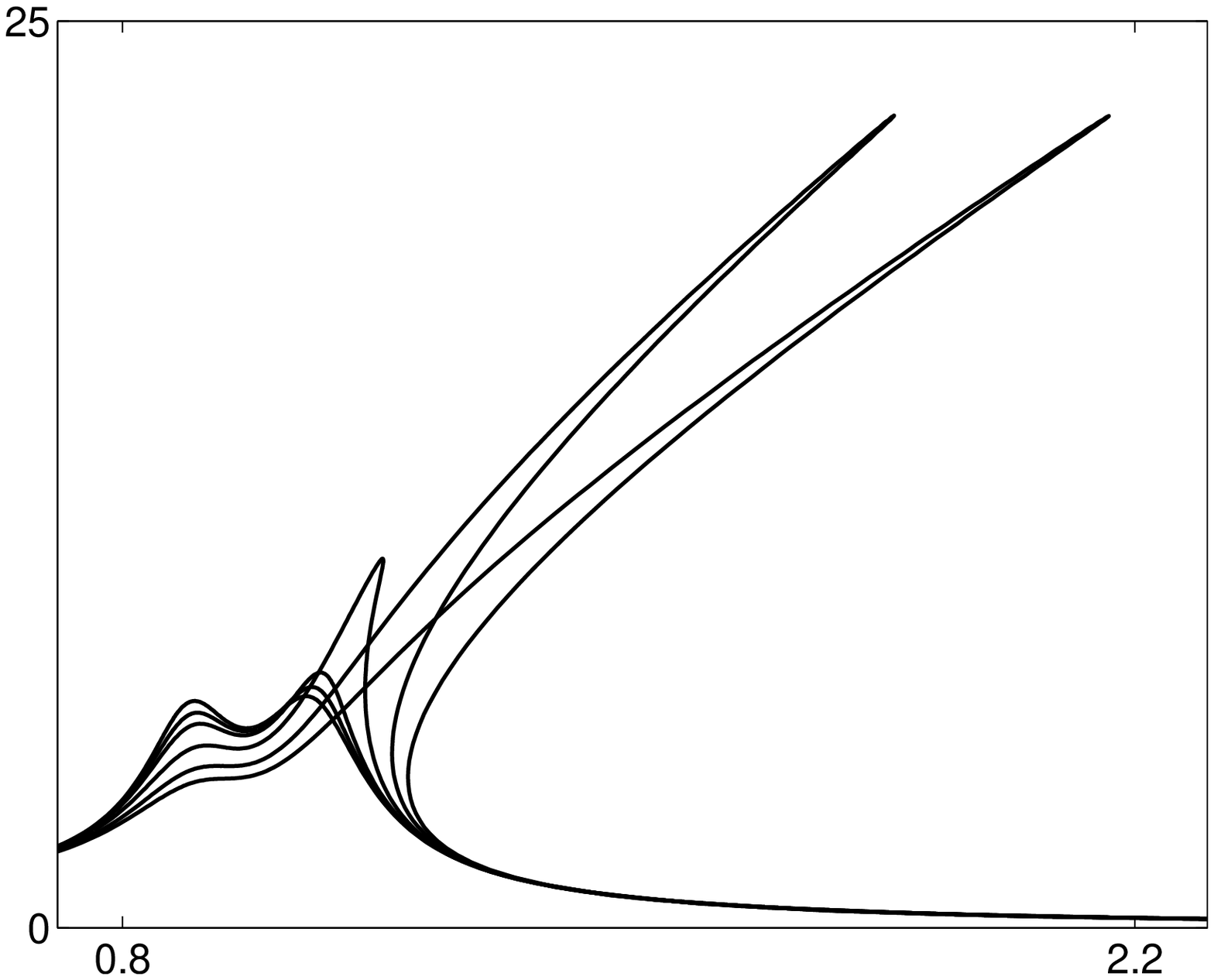}}
\put(3.9,0.4){$\gamma$} \put(12.2,0.4){$\gamma$}  \put(0,3.4){\rotatebox{90}{$q_1$}} \put(8.5,3.4){\rotatebox{90}{$q_1$}}

\put(0.8,5.6){(a)} \put(9.3,5.6){(b)} 
\end{picture}\caption{Numerical solution of Equations (\ref{EqAdim3}) for $\epsilon=0.05$, $\mu_1=0.001$, $\mu_2=0.134$, $\lambda=0.952$. 
Curves from left to right: $\alpha_3=0.0001, 0.0005, 0.001, 0.0025, 0.005, 0.0075$. (a) NLTVA; (b) LTVA. }\label{AdimNEP}
\end{figure}

The outcome of this numerical procedure is displayed in Figure \ref{NL_DH}(a). This plot is  
interesting, because $\beta_3$ is almost linearly related to $\alpha_3$ for 
the different mass ratios considered, i.e., $\beta_3\cong a\alpha_3$. This linear relation implies that the nonlinear coefficient of the 
NLTVA that realizes equal peaks does not depend on forcing amplitude:
\begin{equation}
  \beta_3\cong a\alpha_3 \rightarrow \frac{3}{4}\frac{f^2g'''(r)|_{r=0}}{3!m_2\omega_{n1}^2}\cong a \frac{3}{4}\frac{f^2k_{nl1}}{k_1} \rightarrow 
  g'''(r)|_{r=0}\cong 6a\epsilon k_{nl1}
\end{equation}
The coefficient $a$ is determined by representing $\beta_3$ in function of $\epsilon$ for different values of $\alpha_3$, as in Figure \ref{NL_DH}(b). 
It turns out that the regression $\beta_3=2\alpha_3\epsilon/(1+4\epsilon)$ provides an excellent approximation to the numerical results; 
so $a=2\epsilon/(1+4\epsilon)$. 

Equations (\ref{EqAdim3}) are now solved considering this analytic expression of $\beta_3$ for different values of $\alpha_3$ and $\gamma$, and 
results are presented in Figure \ref{AdimNEP}. We stress that these results were not computed using the one-term harmonic balance approximation but rather 
using the path-following algorithm mentioned in Section \ref{LDO}. This algorithm provides a very accurate numerical solution to the equations of motion. 
Figure \ref{AdimNEP}(a) shows that the NLTVA can enforce equal peaks in the frequency response $q_1$ of the Duffing oscillator for values of $\alpha_3$ ranging 
from $0.0001$ to $0.0075$. This result is remarkable in view of the variation of the resonance frequencies. For instance, 
the first resonance peak occurs at $\gamma=0.9$ for $\alpha_3=0.0001$ and beyond $\gamma=1$ for $\alpha_3=0.0075$. Another interesting observation 
is that the amplitude of the resonance peaks does not change substantially when $\alpha_3$ increases, which means that the response of the coupled system is almost proportional to the forcing amplitude, 
as it would be the case for a linear system. Conversely, Figure \ref{AdimNEP}(b) illustrates that the LTVA is strongly detuned. 
All these results confirm the efficacy of the proposed NLTVA design. 

In summary, given $m_1$, $c_1$, $k_1$ and $k_{nl1}$ for a Duffing oscillator and given a mass ratio $\epsilon$, 
the NLTVA parameters can be determined using the following analytic formulas:
\begin{eqnarray}\label{AnalyticFormulas}
   \nonumber m_2&=&\epsilon m_1 \\
   \nonumber k_2&=&\displaystyle\frac{8\epsilon k_1\left[16+23\epsilon+9\epsilon^2+2(2+\epsilon)\sqrt{4+3\epsilon} \right]}
    {3(1+\epsilon)^2(64+80\epsilon+27\epsilon^2)}\\
   \nonumber c_2&=&\displaystyle\sqrt{\frac{k_2m_2(8+9\epsilon-4\sqrt{4+3\epsilon})}{4(1+\epsilon)}}\\
   k_{nl2}&=&\displaystyle\frac{2\epsilon^2k_{nl1}}{(1+4\epsilon)}
  \end{eqnarray} 
These formulas form the basis of a new tuning rule for nonlinear absorbers that may be viewed as a nonlinear generalization of Den Hartog's equal-peak method. 
We note, however, that there are no invariant points in the nonlinear case. There is thus no complete equivalence with the linear equal-peak method. 
 
\section{Performance of the nonlinear tuned vibration absorber}

\begin{figure}[p]
\setlength{\unitlength}{1cm}
\begin{picture}(8,6.5)(0,0)
\put(0.4,0.6){\includegraphics[width=7.4truecm]{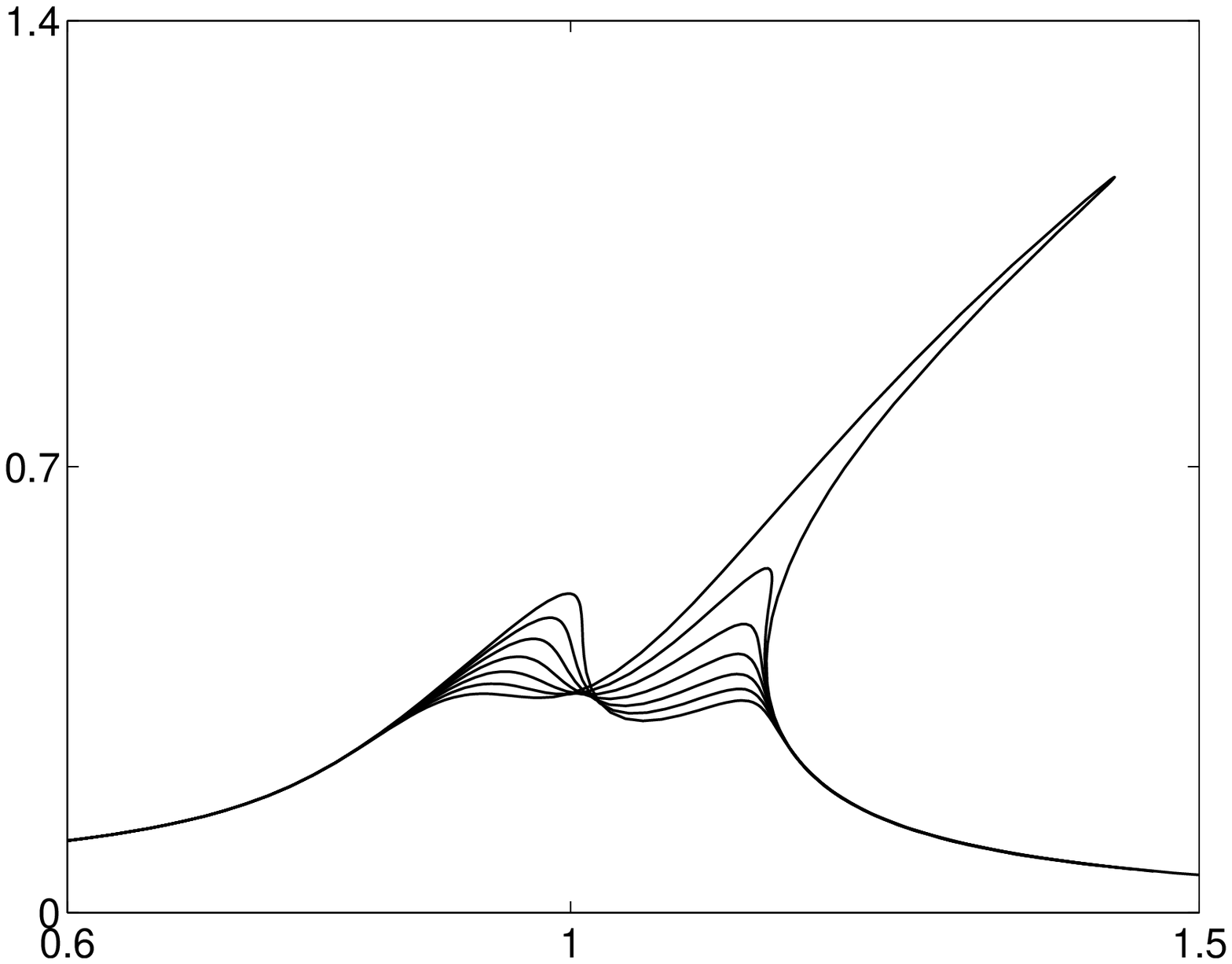}}
\put(8.7,0.6){\includegraphics[width=7.3truecm]{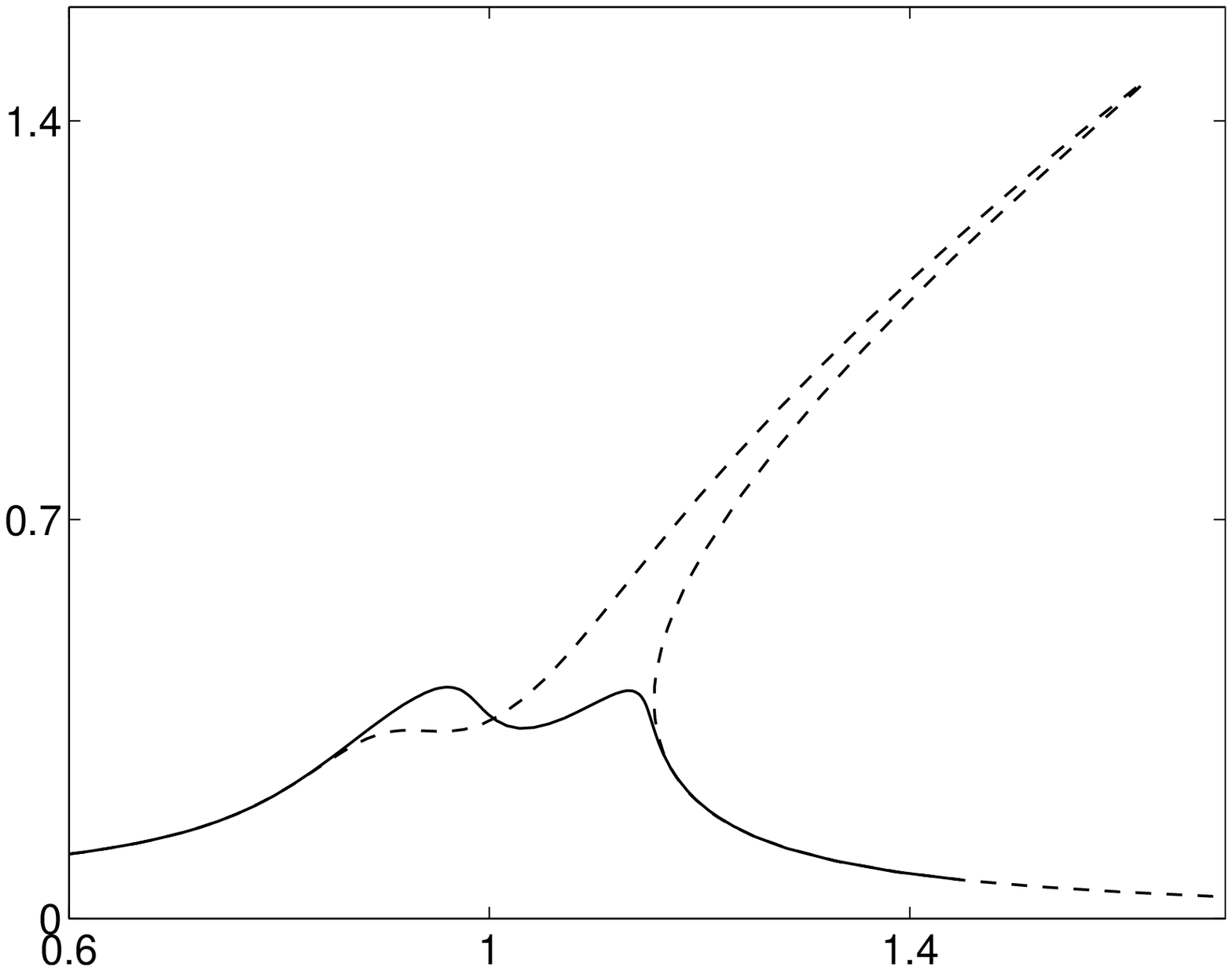}}
\put(2.6,0){Frequency (rad/s)} \put(10.9,0){Frequency (rad/s)}  \put(-0.2,1.8){\rotatebox{90}{Displacement
(m)}} \put(8.1,1.8){\rotatebox{90}{Displacement
(m)}}
\put(1.3,1){\small{0.001$\,$N/m$^3$}}
\put(2.7,1.4){\vector(1,1){0.6}}
\put(4,5.1){\small{0.001$\,$N/m$^3$}}\put(5.7,4.9){\vector(1,-1){0.4}}
\put(1.6,3.3){\small{0.007$\,$N/m$^3$}}\put(3.35,3.2){\vector(1,-1){0.4}}
\put(3.65,1){\small{0.007$\,$N/m$^3$}}\put(4.2,1.4){\vector(1,1){0.6}}
 \put(1.2,5.6){(a)} \put(9.7,2.8){NLTVA} \put(10.6,2.7){\vector(1,-1){0.5}} \put(13.6,5.25){\vector(1,-1){0.5}}
  \put(13,5.4){LTVA} \put(9.6,5.6){(b)} 
\end{picture}\caption{Frequency response of a Duffing oscillator with an attached NLTVA for $F=0.07\,$N for (a) various nonlinear NLTVA coefficients $k_{nl2}$; (b) $k_{nl2}=0.0042\,$N/m$^3$.}\label{NonlinearEqualPeakWellTuned}
\end{figure}

\begin{figure}[p]
\setlength{\unitlength}{1cm}
\begin{picture}(8,13.5)(0,0)
\put(0.75,7){\includegraphics[width=7.3truecm]{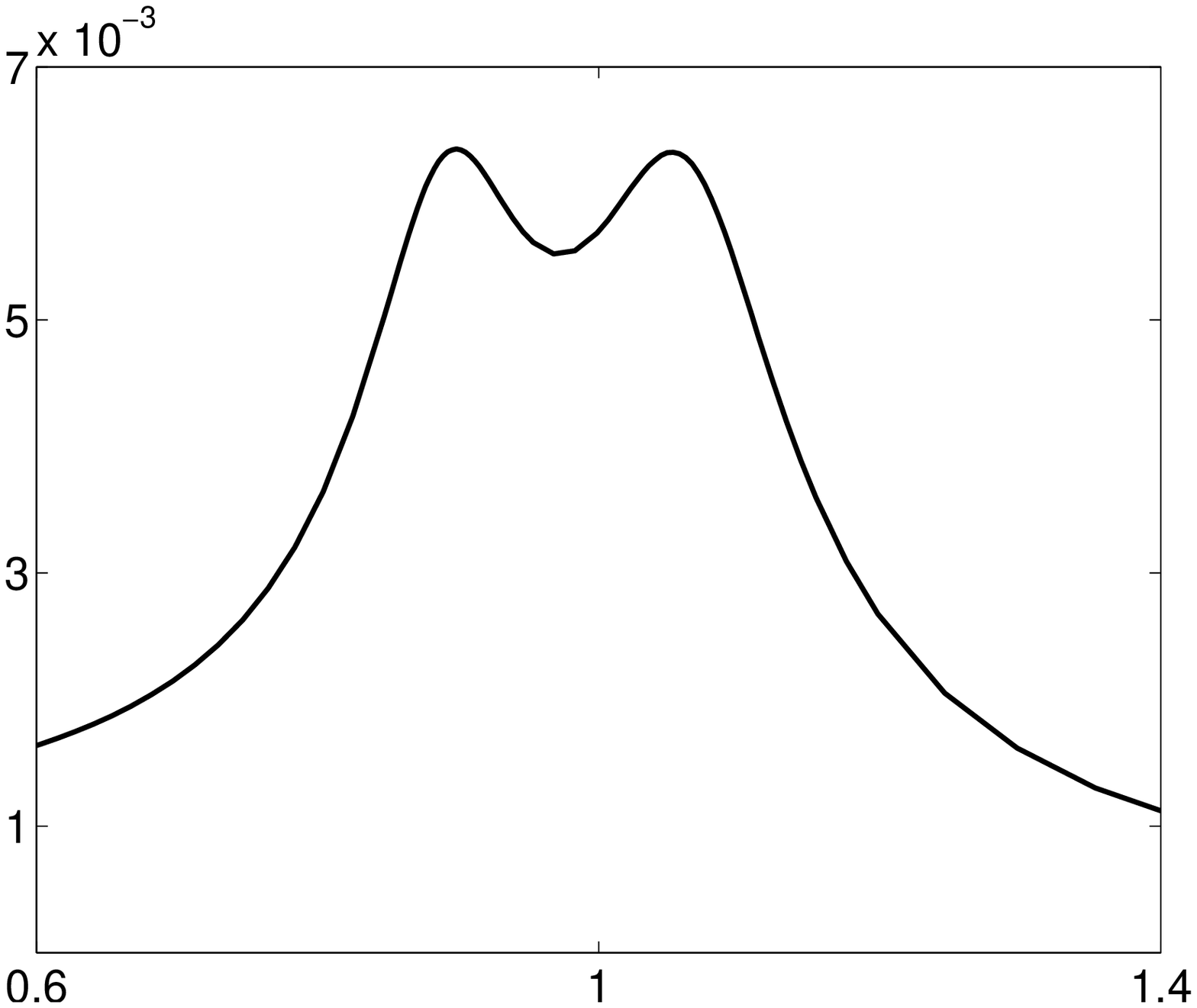}}
\put(8.45,7){\includegraphics[width=7.5truecm,height=5.9truecm]{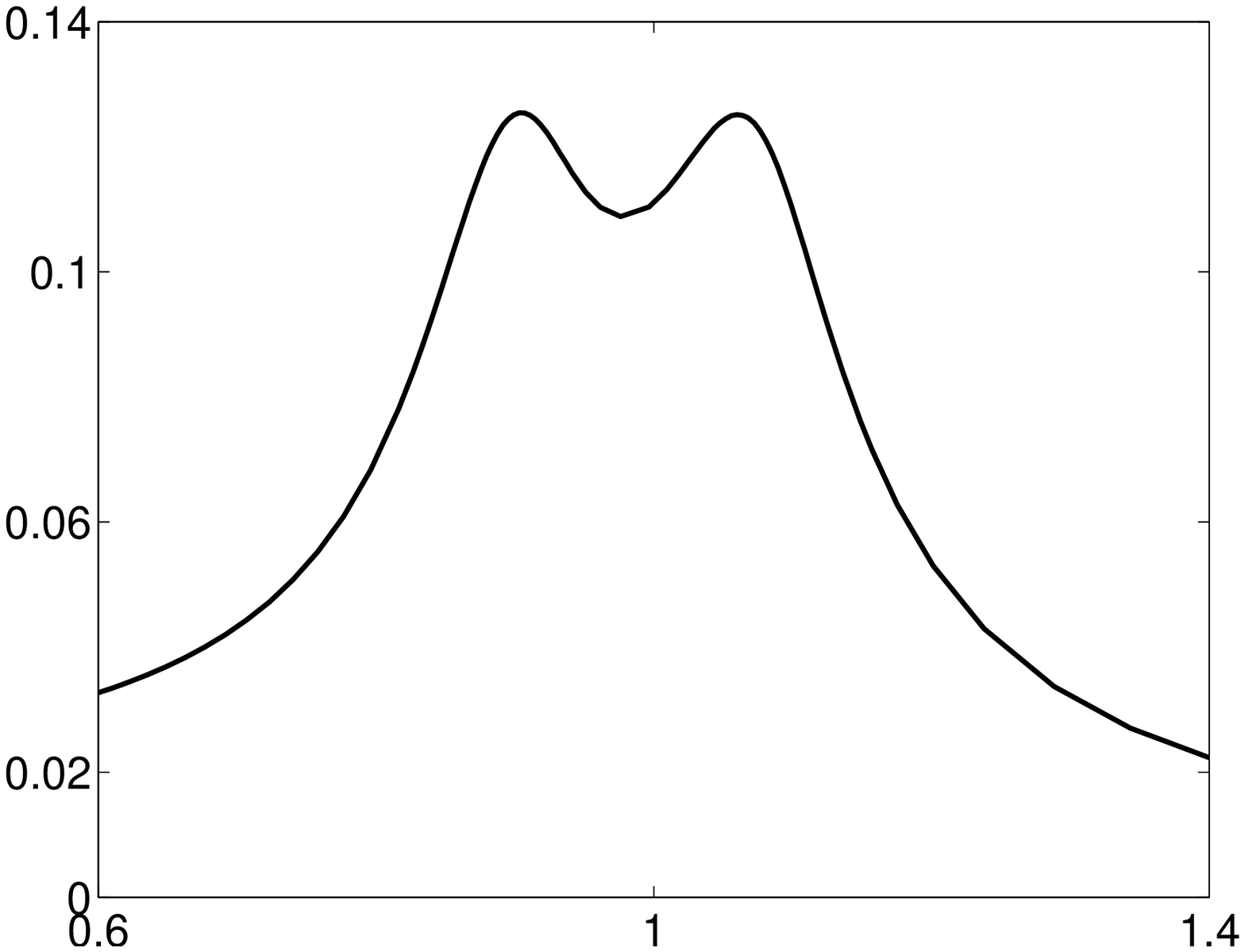}}
\put(0.5,0.6){\includegraphics[width=7.5truecm]{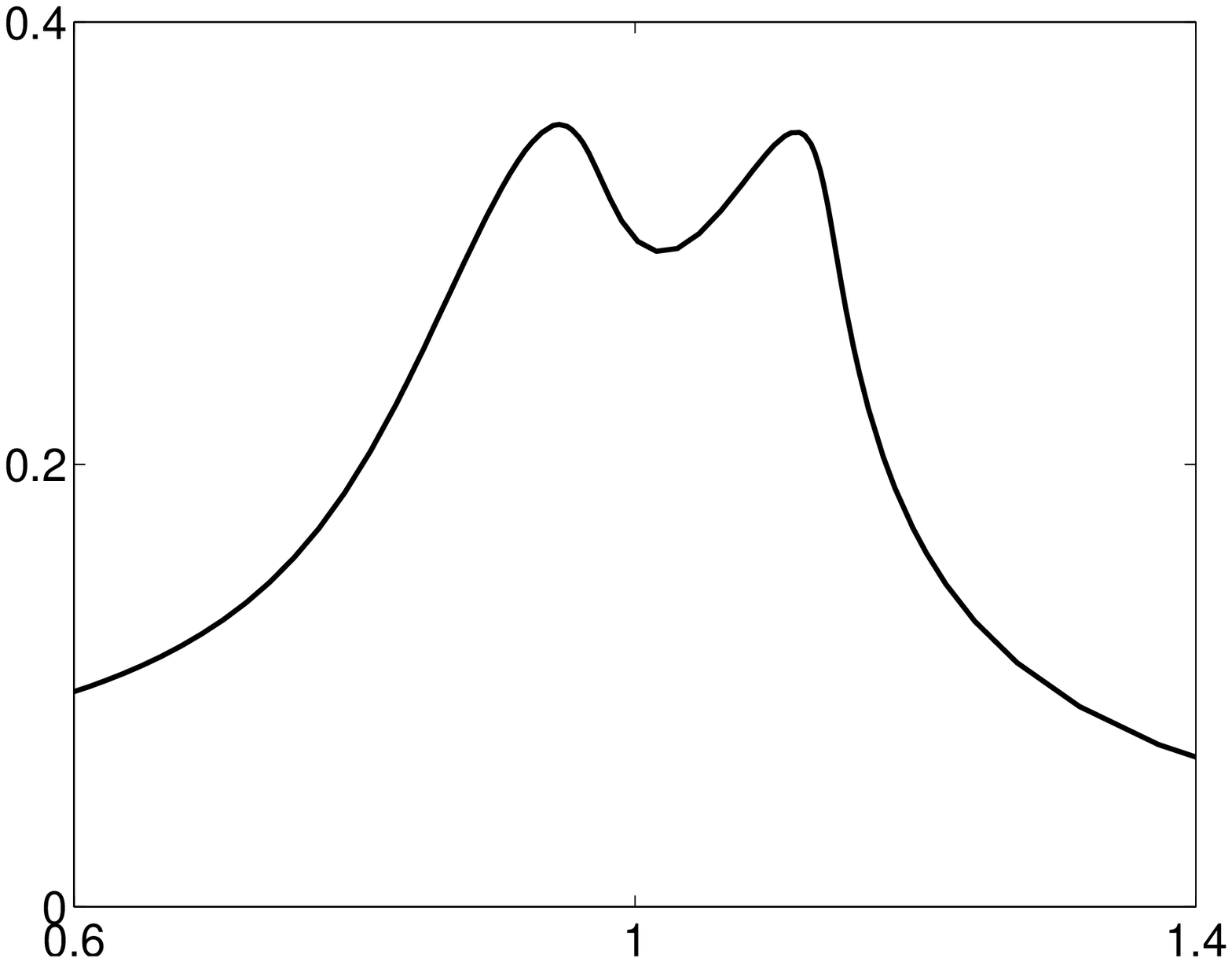}}
\put(8.5,0.6){\includegraphics[width=7.5truecm]{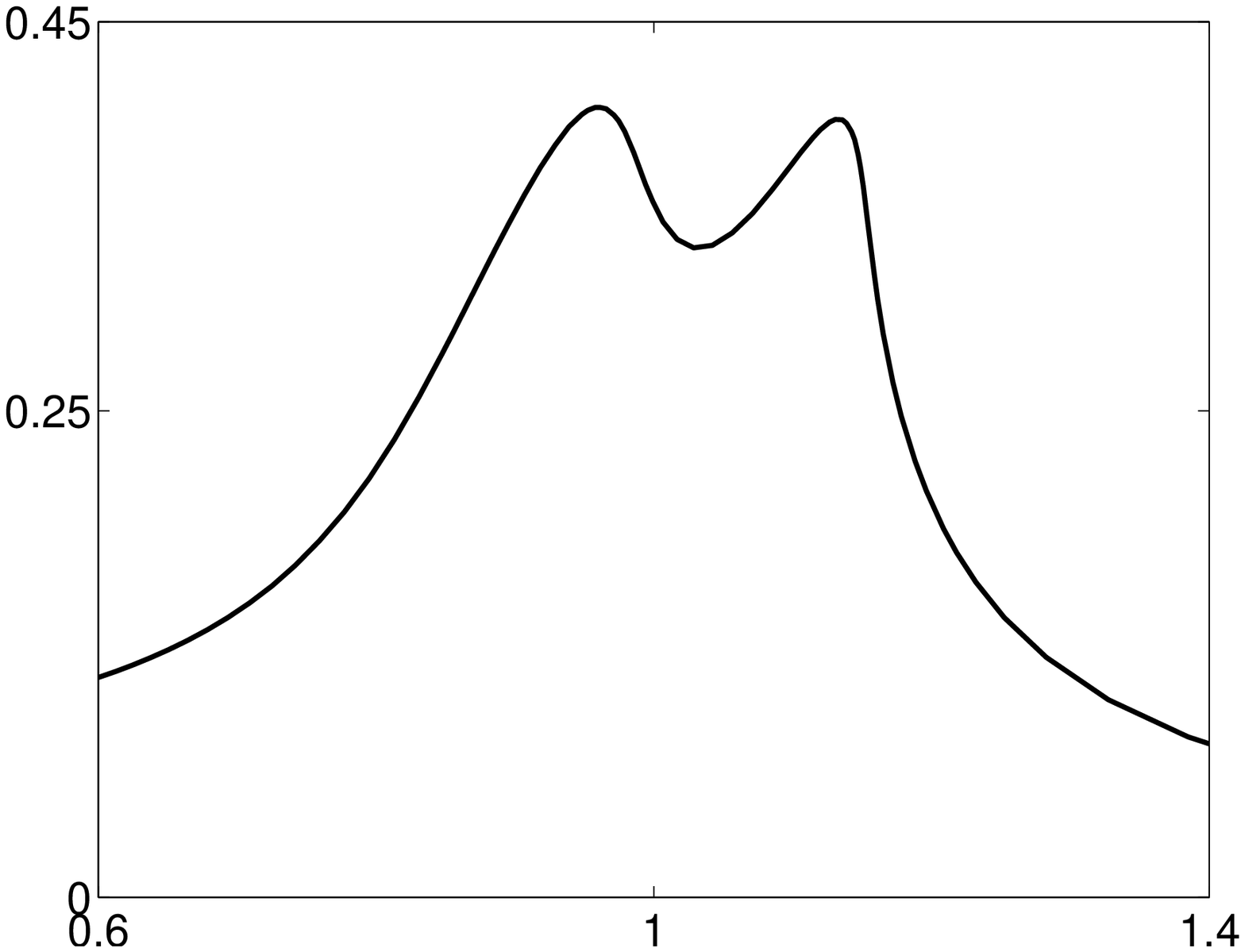}}
\put(6.6,0){Frequency (rad/s)} \put(-0.1,5.3){\rotatebox{90}{Displacement
(m)}}
 \put(1.2,12){(a)}
 \put(9.3,12){(b)}
 \put(1.2,5.7){(c)}
 \put(9.3,5.7){(d)}
\end{picture}\caption{Frequency response of a Duffing oscillator with an attached NLTVA. (a) $F=0.001\,$N; (b) $F=0.02\,$N; (c) $F=0.06\,$N, and (d) $F=0.07\,$N.}\label{PerfoNLTVA}
\end{figure}

The previous theoretical developments are further validated and illustrated in the present section. An NLTVA possessing 
a linear and a cubic spring is attached to a Duffing oscillator with $m_1=1\,$kg, $c_1=0.002\,$Ns/m, $k_1=1\,$N/m and $k_{nl1}=1\,$N/m$^3$. 
The mass ratio is 5\%. According to Equations (\ref{AnalyticFormulas}), the NLTVA parameters should be $m_2=0.05\,$kg, $c_2=0.0128\,$Ns/m, 
$k_2=0.0454\,$N/m and $k_{nl2}=0.0042\,$N/m$^3$. Figure \ref{NonlinearEqualPeakWellTuned}(a) presents the displacement of the primary system 
for a forcing amplitude $F=0.07\,$N and for nonlinear coefficients $k_{nl2}$ between 0.001 to 0.007$\,$N/m$^3$. As illustrated in Figure 
\ref{NonlinearEqualPeakWellTuned}(b), resonance peaks of equal amplitude are obtained when $k_{nl2}=0.0042\,$N/m$^3$. 
The response with an attached LTVA is also superposed and confirms the substantial improvement brought by the NLTVA. Figure \ref{PerfoNLTVA} 
plots the displacement of the Duffing oscillator for $F$ ranging from 0.001$\,$N to 0.07$\,$N and for $k_{nl2}=0.0042\,$N/m$^3$. This figure is to be 
compared with Figure \ref{LINTVA}.

\begin{figure}[p]
\setlength{\unitlength}{1cm}
\begin{picture}(8,9)(0,0)
\put(0.5,0.6){\includegraphics[width=7.2truecm]{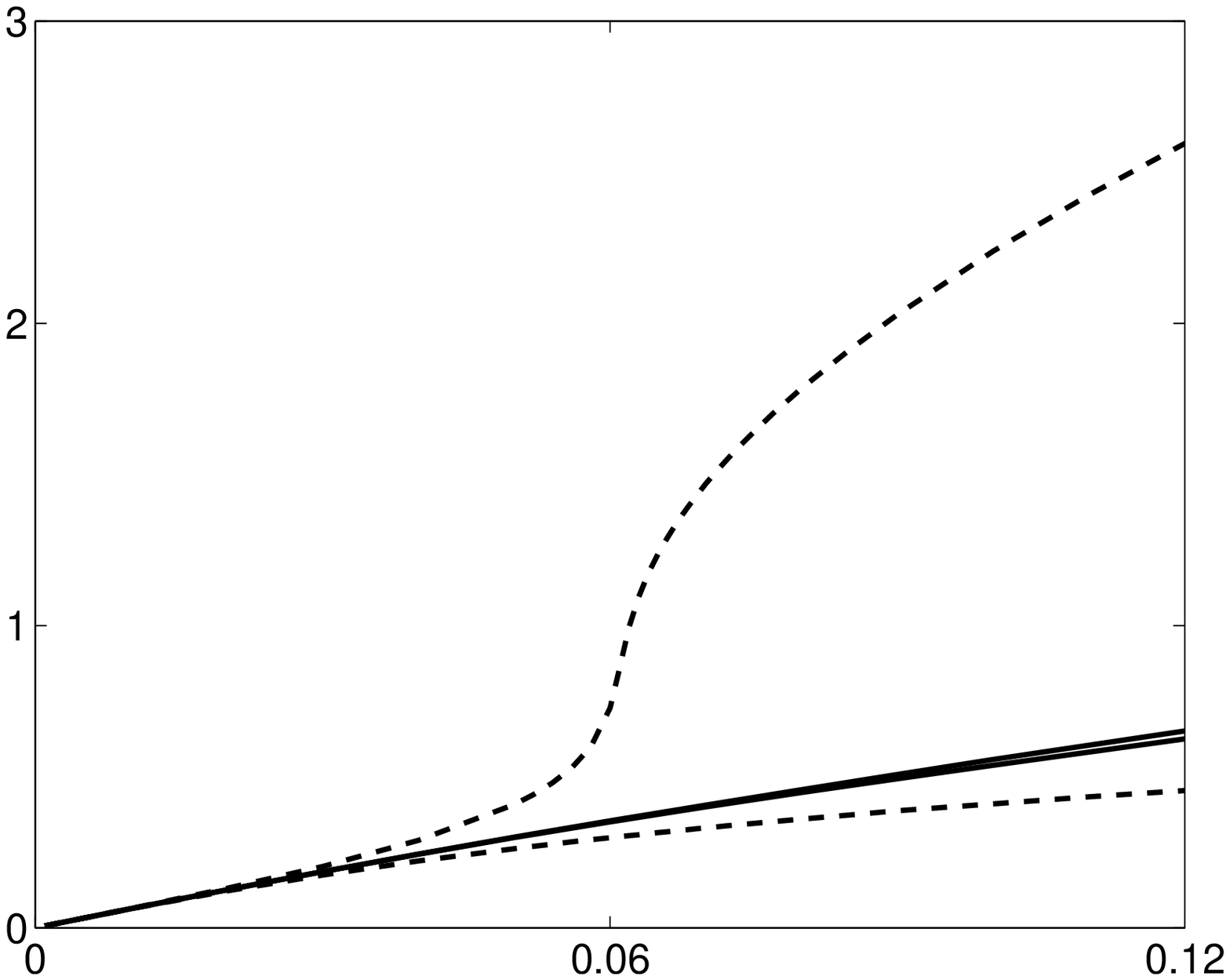}}
\put(8.7,0.6){\includegraphics[width=7.4truecm]{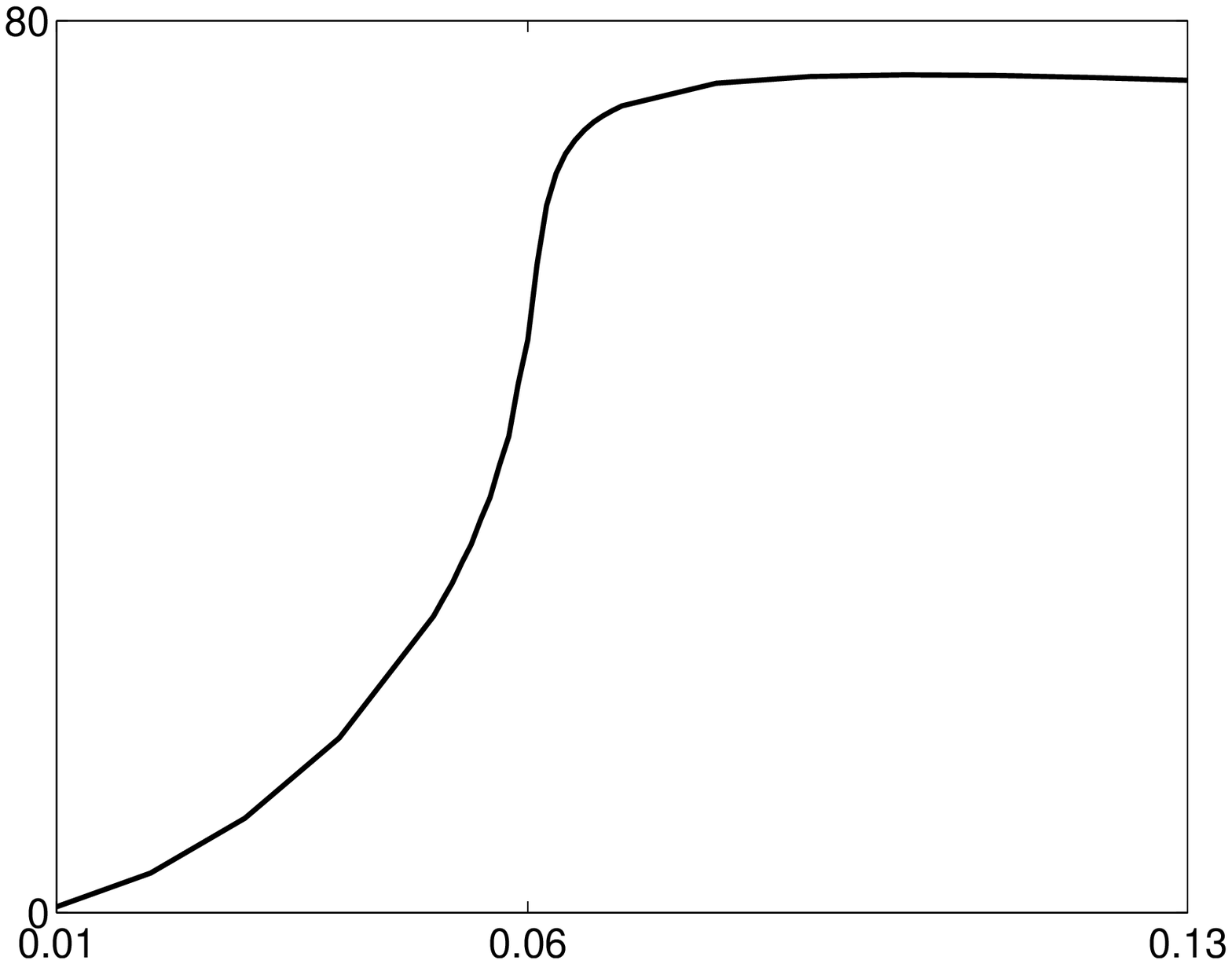}}
\put(6.1,0.1){Forcing amplitude (N)} \put(-0.2,1.995){\rotatebox{90}{Displacement
(m)}} \put(8.2,1.2){\rotatebox{90}{NLTVA improvement(\%)}}
 \put(1.2,5.7){(a)}
 \put(9.3,5.7){(b)}
\end{picture}\caption{Performance of the NLVTA/LTVA for increasing forcing amplitudes. (a) Amplitude of the resonances peaks of the Duffing oscillator 
(solid lines: NLTVA, dashed lines: LTVA); (b) percentage of improvement brought by the NLTVA with respect to the LTVA.}\label{CompaNLTVA_LTVA_closeup}
\end{figure}

%\begin{figure}[p]
%\setlength{\unitlength}{1cm}
%\begin{picture}(8,9)(0,0)
%\put(0.5,0.6){\psfig{figure=CompaNLTVA_LTVA_large.eps,width=7.5truecm}}
%\put(8.7,0.6){\psfig{figure=CompaNLTVA_LTVA_isola.eps,width=7.5truecm}}
%\put(1.2,5.7){(a)}
% \put(9.3,5.7){(b)}
%\put(6.3,0){Forcing amplitude (N)}   \put(-0.2,1.995){\rotatebox{90}{Displacement (m)}} \put(8,1.995){\rotatebox{90}{Displacement (m)}}
% \end{picture}\caption{Peak amplitude of the Duffing oscillator (solid lines: NLTVA, dashed lines: LTVA). (a) Without the detached resonance curve; (b) with the detached resonance curve.}\label{CompaNLTVA_LTVA_large}
%\end{figure}

\begin{figure}[p]
\setlength{\unitlength}{1cm}
\begin{picture}(8,9)(0,0)
\put(4.5,0.6){\includegraphics[width=7.5truecm]{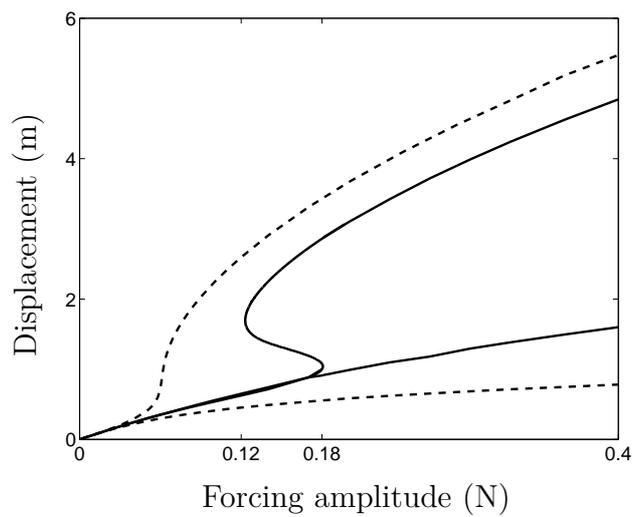}}
\put(6.3,0){Forcing amplitude (N)}   \put(3.8,1.995){\rotatebox{90}{Displacement (m)}} 
 \end{picture}\caption{Peak amplitude of the Duffing oscillator (solid lines: NLTVA, dashed lines: LTVA).}\label{CompaNLTVA_LTVA_large}
\end{figure}

\begin{figure}[t]
\setlength{\unitlength}{1cm}
\begin{picture}(8,7)(0,0)
\put(0.5,0.6){\includegraphics[width=7.4truecm]{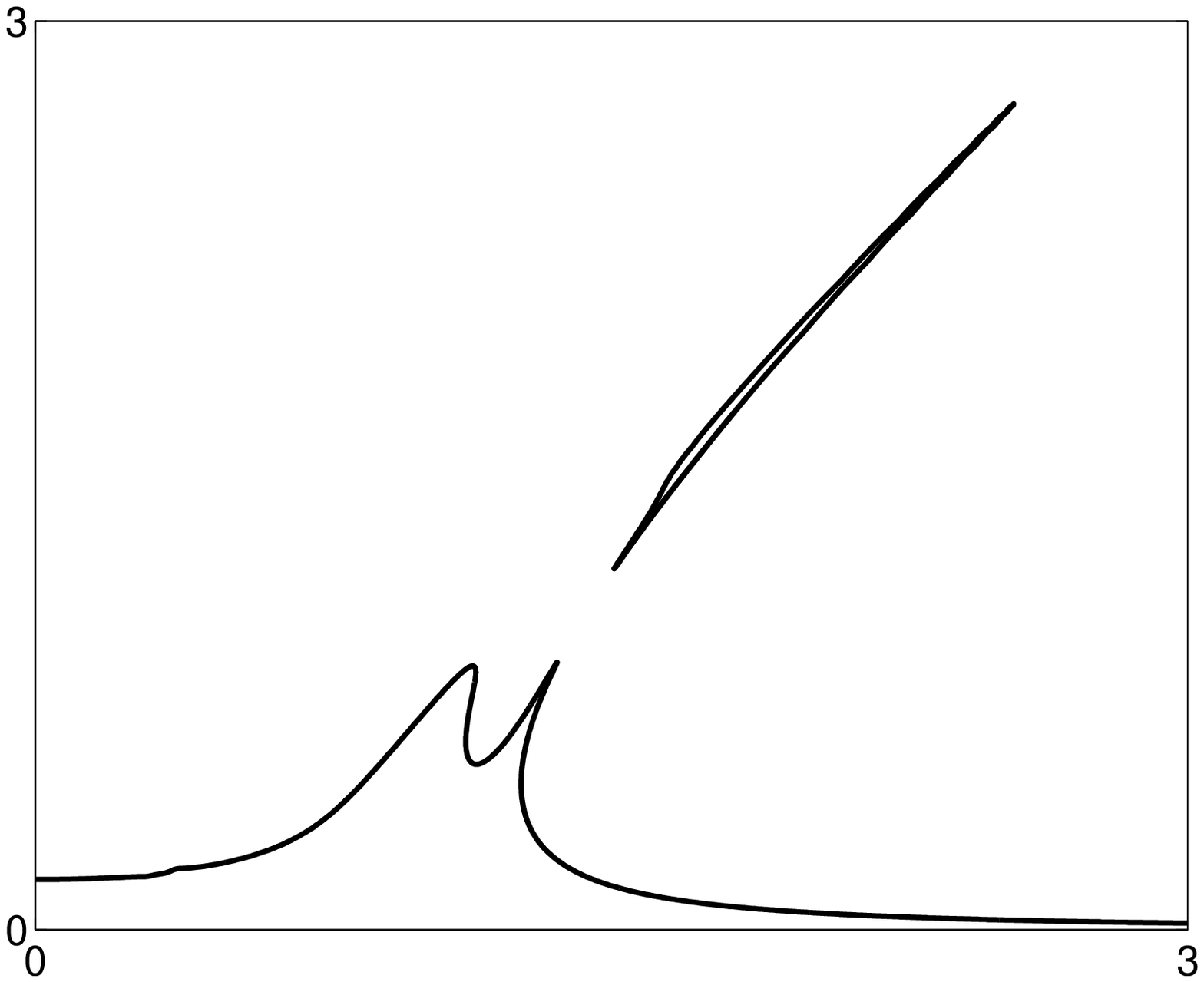}}
\put(8.7,0.6){\includegraphics[width=7.4truecm]{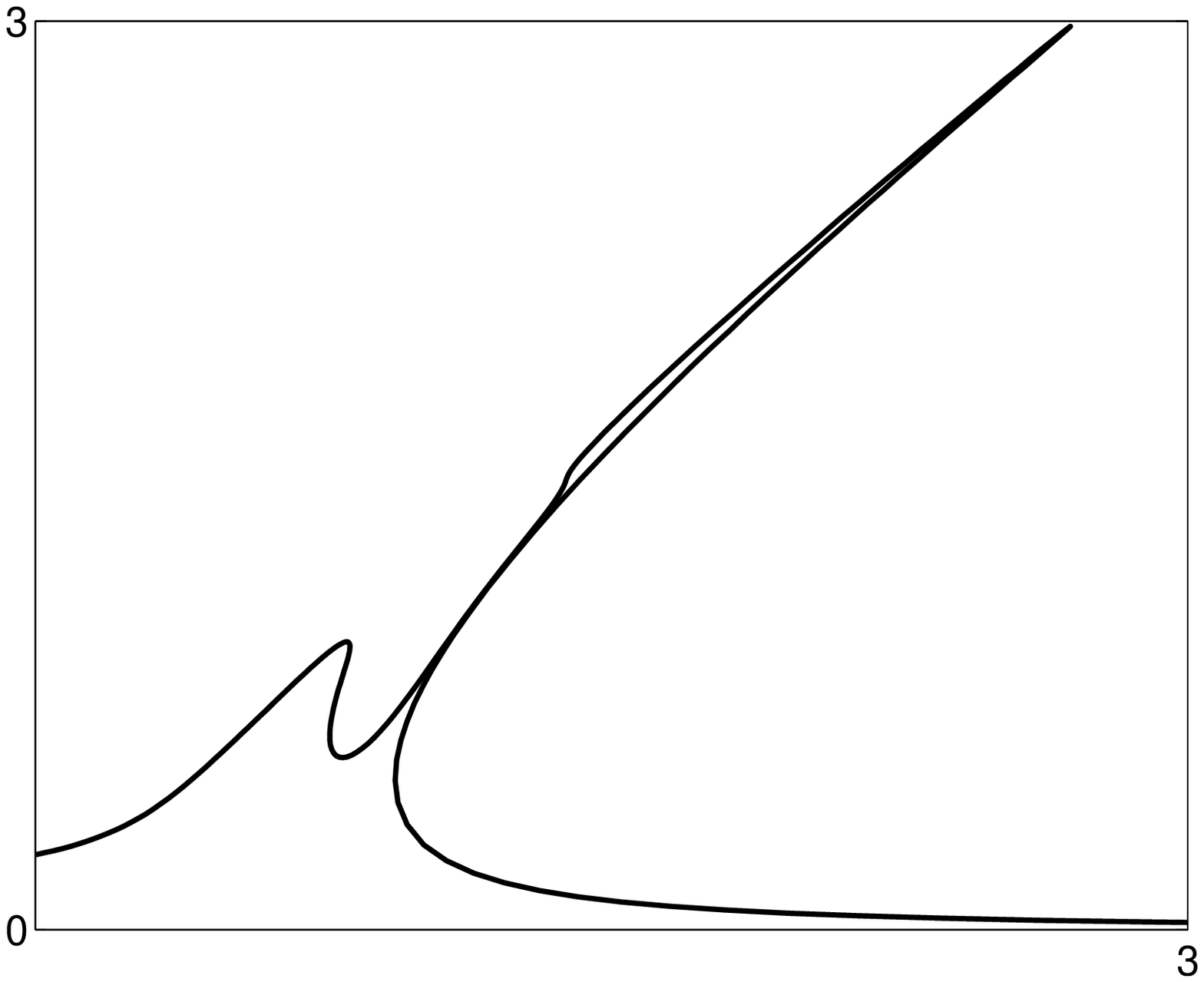}}
\put(6.3,0.1){Frequency (rad/s)} \put(0,1.995){\rotatebox{90}{Displacement
(m)}} \put(8.2,1.995){\rotatebox{90}{Displacement (m)}}
 \put(1.05,5.7){(a)}
 \put(9.3,5.7){(b)}
\end{picture}\caption{Detuning of the NLTVA. (a) $F=0.17\,$N: presence of a detached resonance curve; (b) $F=0.19\,$N: the detached resonance curve has merged 
with the largest resonance peak.}\label{ExplanationDetuning}
\end{figure}

For a more global comparison between the two absorbers, Figure \ref{CompaNLTVA_LTVA_closeup}(a) represents the amplitude of the resonance peaks of the 
Duffing oscillator as a function of $F$. If the LTVA gets rapidly detuned, the amplitude of the two resonance peaks for the NLTVA remains almost identical. 
In addition, the amplitude is almost linearly related to forcing amplitude, as if the system would obey the superposition principle. This result is unexpected 
in view of the strongly nonlinear regimes investigated. It therefore seems that adding a properly chosen nonlinearity to an already nonlinear system can somehow 
linearize the dynamics of the coupled system. On the contrary, the amplitude of the resonance peaks for the LTVA exhibits a marked nonlinear dependence with respect 
to forcing amplitude. Figure \ref{CompaNLTVA_LTVA_closeup}(b) illustrates that the NLTVA performance is always superior to that of 
the LTVA, which is a further appealing feature of this device.

If the dynamics is investigated for even larger amplitudes, an important detuning of the NLTVA occurs in 
Figure \ref{CompaNLTVA_LTVA_large}. This detuning can be explained by the presence of a detached resonance curve, also termed an isola, 
which is similar in essence to that reported for the LTVA in Section \ref{LDO}. For $F=0.17\,$N, Figure \ref{ExplanationDetuning}(a) shows that equal peaks 
are still maintained, but they co-exist with the isola. The characterization of the isola
was carried out numerically using a co-dimension 2 bifurcation tracking procedure based on the harmonic balance method. The description of this methodology 
is beyond the scope of this paper. The interested reader may refer to reference \cite{Detroux} for further details.

\begin{figure}[p]
\setlength{\unitlength}{1cm}
\begin{picture}(8,6.2)(0,0)
\put(4.5,0.6){\includegraphics[width=7.4truecm]{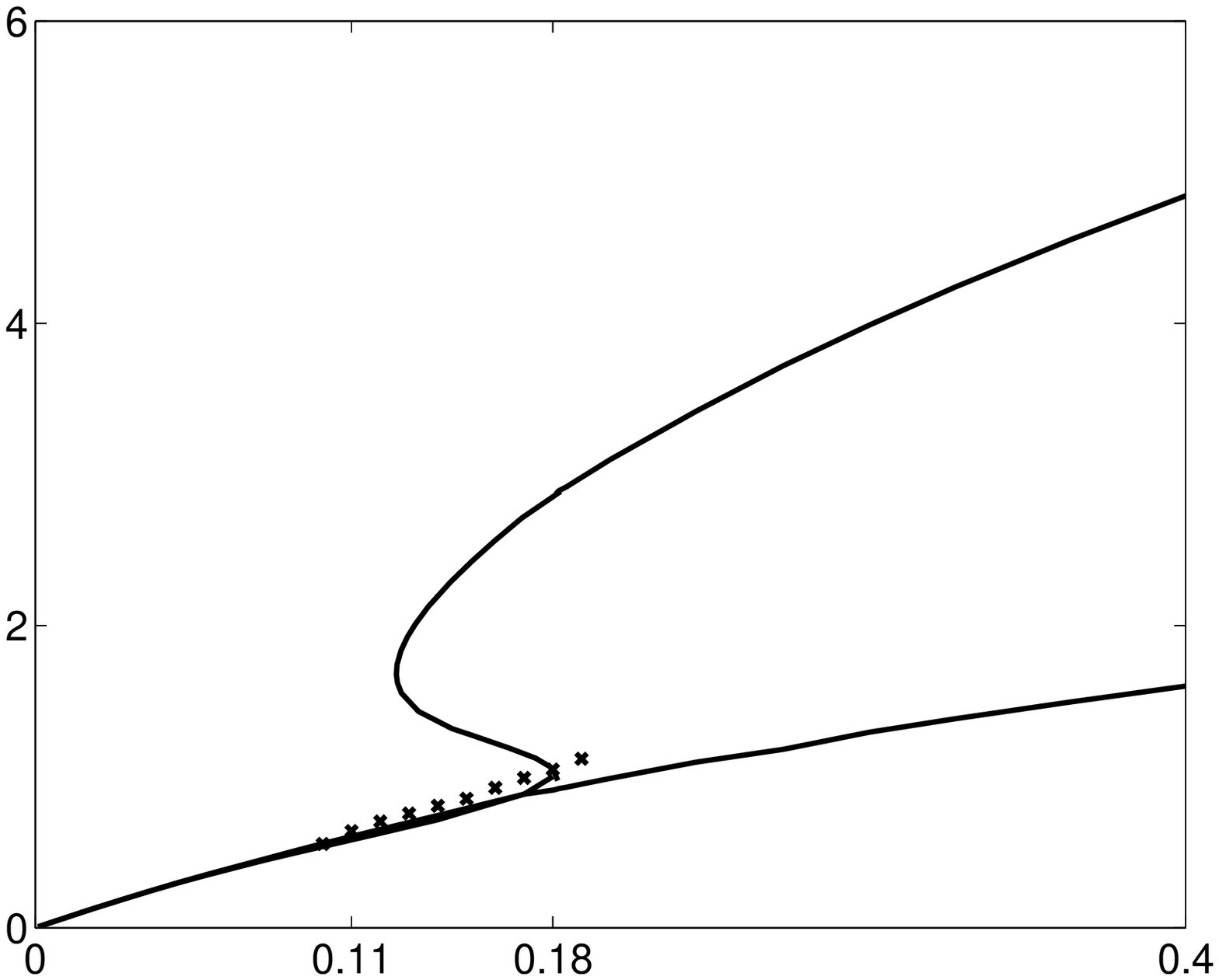}}
\put(6.1,0.1){Forcing amplitude (N)} \put(3.8,1.995){\rotatebox{90}{Displacement
(m)}} \end{picture}\caption{Peak amplitude of the Duffing oscillator with an attached NLTVA; the quasiperiodic motion of the NLTVA is represented with black crosses.}\label{QP}
\end{figure}
\begin{figure}[p]
\setlength{\unitlength}{1cm}
\begin{picture}(8,6.2)(0,0)
\put(0.5,0.6){\includegraphics[width=7.4truecm]{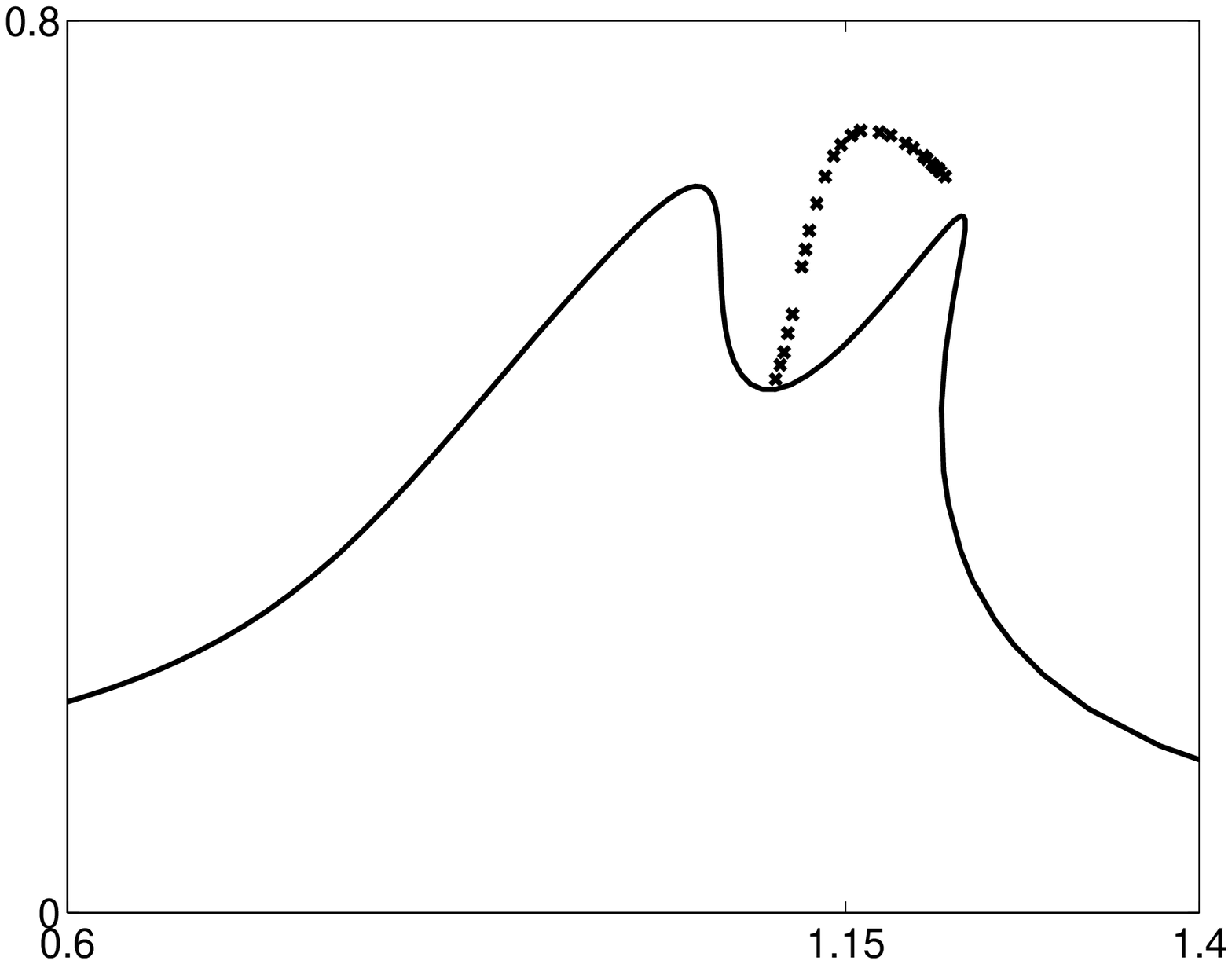}}
\put(8.7,0.6){\includegraphics[width=7.7truecm]{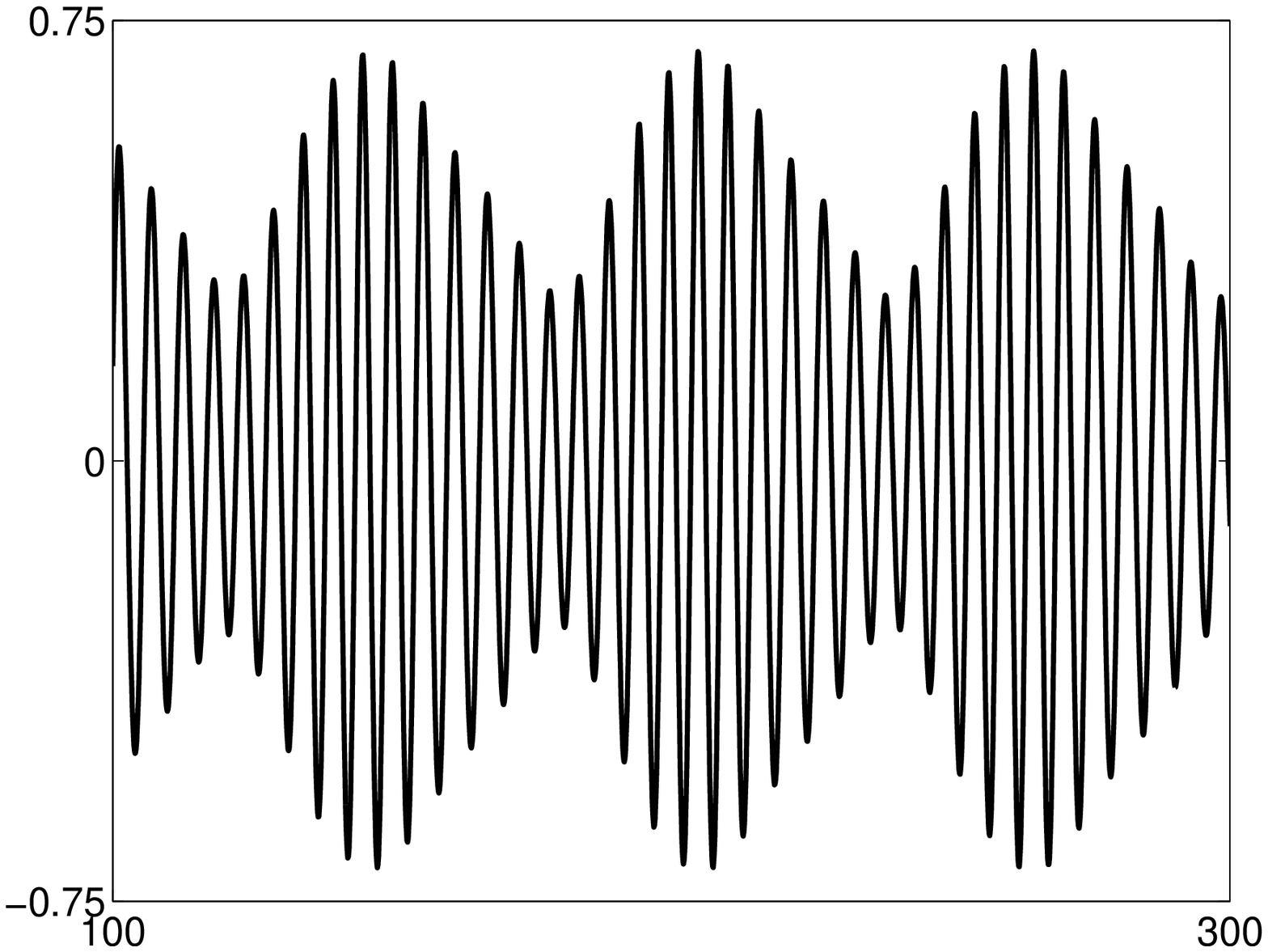}}
\put(11.9,0.25){Time (s)} \put(0.2,1.995){\rotatebox{90}{Displacement
(m)}} \put(2.7,0.25){Frequency (rad/s)} \put(8.5,1.995){\rotatebox{90}{Displacement
(m)}}
 \put(1.2,5.7){(a)}
 \put(9.45,5.7){(b)}
\end{picture}\caption{Quasiperiodic motion of the NLTVA for $F=0.12\,$N. (a) Frequency response of the Duffing oscillator; 
(b) displacement of the Duffing oscillator for $\omega=1.15\,$rad/s.}\label{QP1}
\end{figure}
\begin{figure}[t]
\setlength{\unitlength}{1cm}
\begin{picture}(8,13.6)(0.3,0)
\put(4.25,7.6){\includegraphics[width=8truecm]{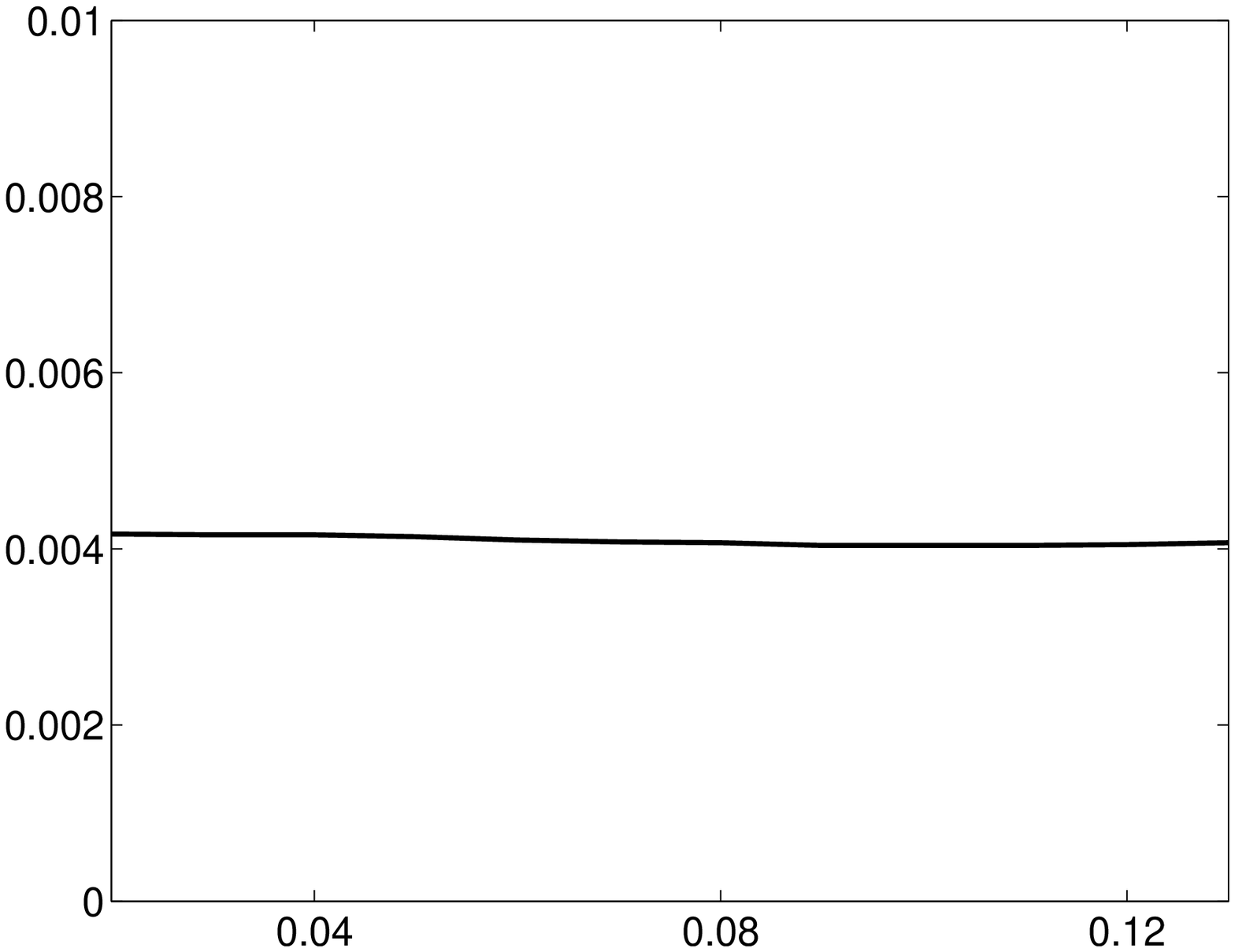}}
 \put(3.65,8.1){\rotatebox{90}{Nonlinear coefficient (N/m$^3$)}}
\put(0.5,0.6){\includegraphics[width=7.5truecm]{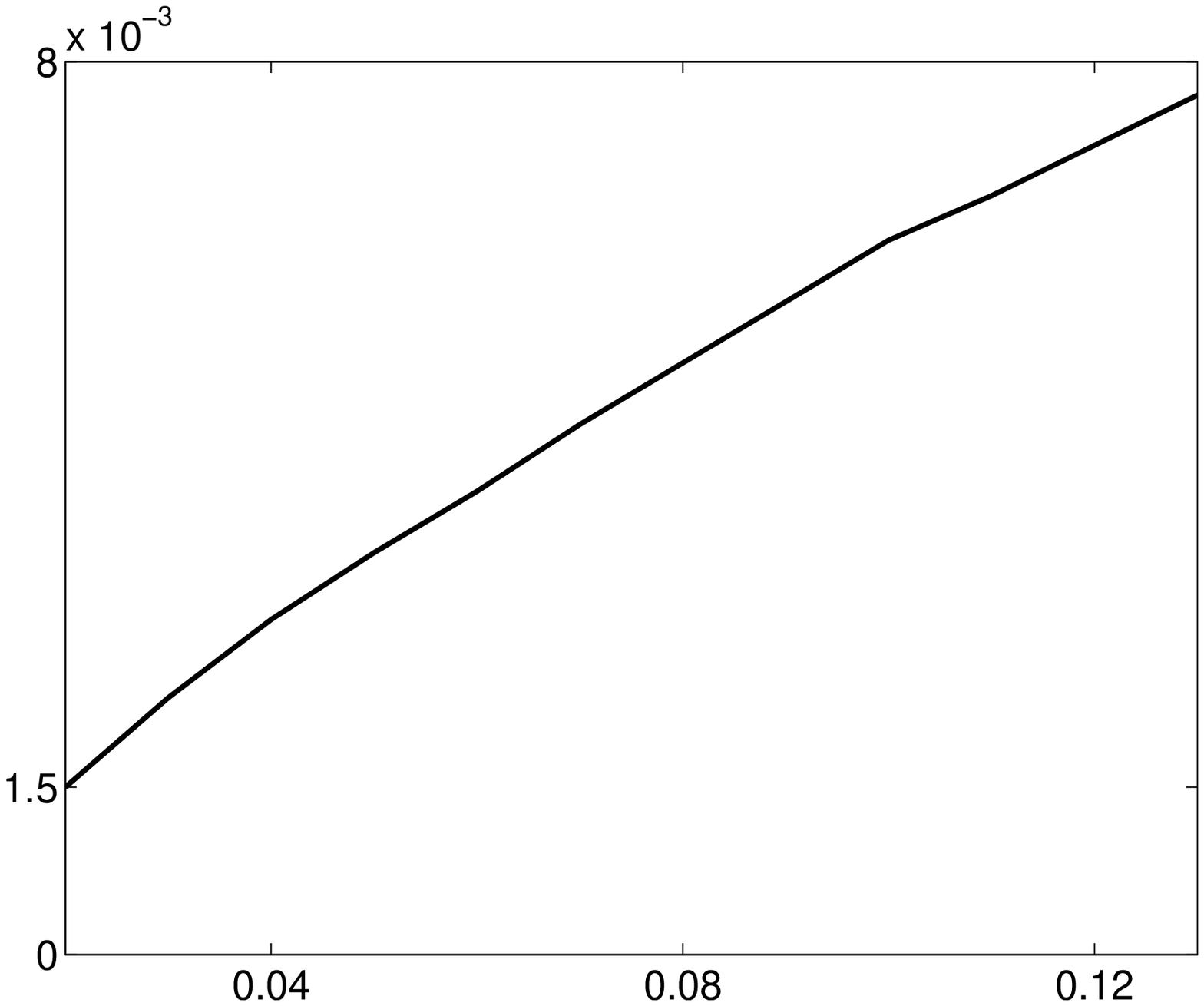}}
\put(8.7,0.6){\includegraphics[width=7.7truecm]{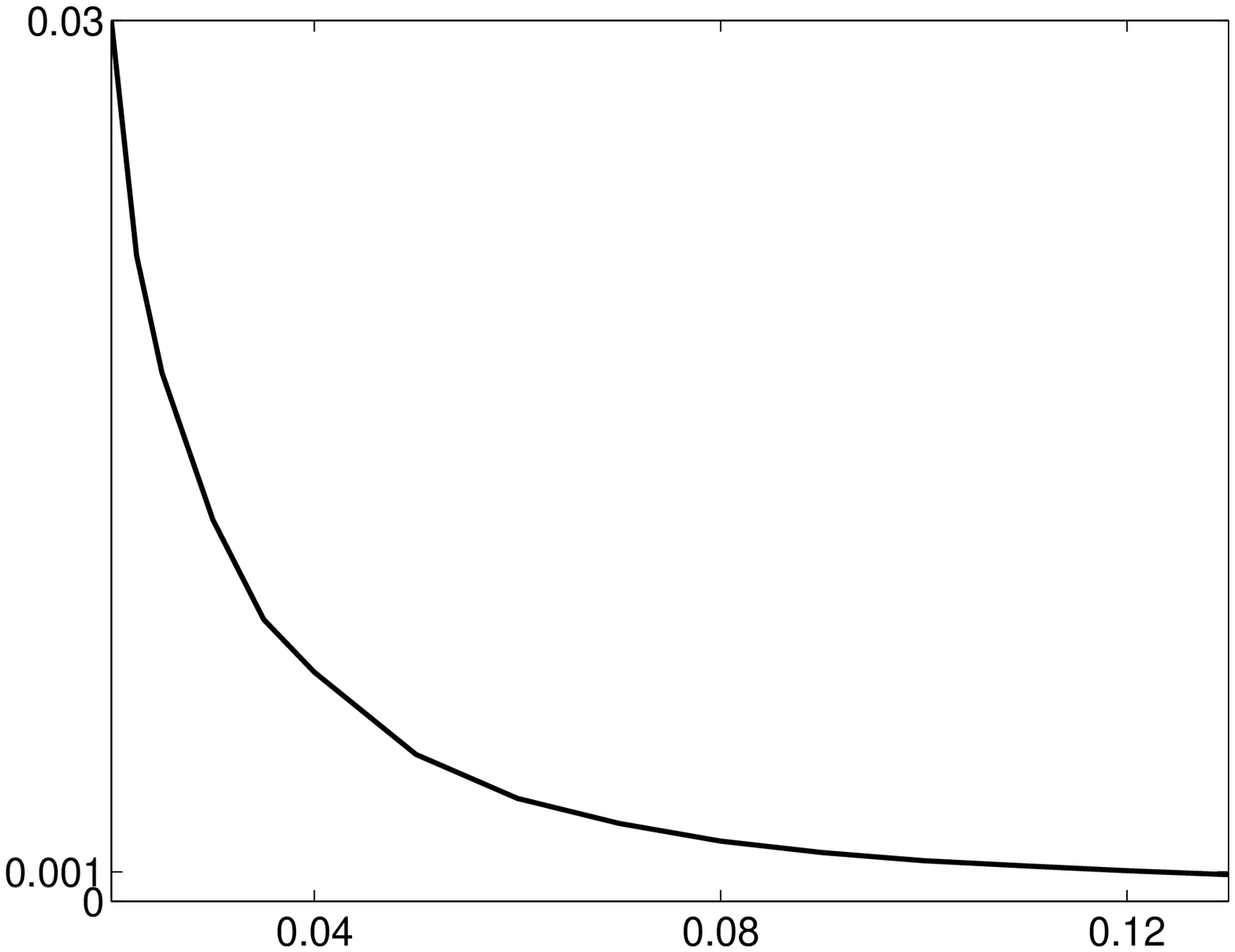}}
\put(6.5,0){Forcing amplitude (N)} \put(6.5,7){Forcing amplitude (N)}  \put(-0.1,1){\rotatebox{90}{Nonlinear coefficient (N/m$^2$)}} 
\put(8.25,1.05){\rotatebox{90}{Nonlinear coefficient (N/m$^5$)}}
\put(5.2,12.8){(a)}
\put(1.2,5.7){(b)}
 \put(9.7,5.7){(c)}
 \end{picture}\caption{Value of $k_{nl2}$ realizing equal peaks at different forcing amplitudes. (a) Cubic spring; (b) quadratic spring; (c) quintic spring.}\label{OtherTuningAmplitudes2}
\end{figure}

The amplitude of the limit points of the isola as a function of forcing amplitude is displayed in Figure \ref{CompaNLTVA_LTVA_large}. The underlying dynamical mechanism is as follows. The isola is created by a pair of saddle-node bifurcations close to $F=0.12\,$N. For increasing forcing amplitudes, 
the two bifurcations move away from each other, as depicted in Figure \ref{CompaNLTVA_LTVA_large}(b), until one of them meets the saddle-node bifurcation defining 
the second resonance peak. The isola then coalesces with this resonance, which gives rise to a new resonance peak with a much larger amplitude, as represented in 
Figure \ref{ExplanationDetuning}(b) for $F=0.19\,$N. The coalescence occurs for $F=0.18\,$N; this forcing amplitude therefore represents an upper limit beyond 
which the nonlinear equal-peak method is no longer applicable. Despite the presence of the isola, we note that the NLTVA still remains 
more effective than the LTVA. 
Furthermore, if the basin of attraction of the isola is relatively small, the absorber can work properly also in coexistence of the isola itself.

Going back to Figures \ref{NL_DH}(a,b), the isola explains why the curves were discontinued, i.e., why equal peaks could no longer be achieved. 
From these plots, we can observe that isolas occur for comparatively larger values of $\alpha_3$ and smaller values of the mass ratio $\epsilon$. 
Isolas are generic for nonlinear vibration absorbers and neutralizers \cite{Brennan1,Brennan3,Staros}. They will be investigated more closely 
in future studies.

Another dynamical instability that exists in the coupled system is a combination resonance leading to quasiperiodic regimes of motions. It appears because the NLTVA is designed such that the desired 
operating frequency is approximately the mean of the two nonlinear resonant frequencies of the system \cite{Shaw4}. Quasiperiodic motion is observed between 
$F=0.10\,$N and $F=0.19\,$N, and the corresponding peak amplitudes are represented in Figure \ref{QP}. For illustration, a quasiperiodic branch computed through direct numerical simulations 
is depicted in Figure \ref{QP1}(a) 
for $F=0.12\,$N, and a time series is shown in Figure \ref{QP1}(b). Unlike what was observed in \cite{Shaw4}, 
the quasiperiodic motions have an amplitude comparable to the amplitude of the resonance peaks. They do not destroy the effectiveness of the NLTVA. We note that, 
for a nonlinear energy sink, quasiperiodic motions can even be beneficial, because they replace a resonance with motions of much smaller amplitudes \cite{Staros}.

To conclude the validation of the nonlinear equal-peak method, the nonlinear coefficient $k_{nl2}$ that realizes equal resonance peaks 
for various forcing amplitudes is computed using the path-following algorithm. Three different nonlinearities for the NLTVA, namely $x^2$, $x^3$ and $x^5$, are considered. 
Figure \ref{OtherTuningAmplitudes2}(a) confirms that the coefficient of the cubic spring does not exhibit variability when the forcing amplitude is increased. 
Its value is in very close agreement with the closed-form solution (\ref{AnalyticFormulas}). For a NLTVA with a quadratic spring, Figure \ref{OtherTuningAmplitudes2}(b)
evidences that the nonlinear coefficient undergoes a strong variation with forcing amplitude, as it was predicted by the discussion in Section \ref{synth}. 
The variation is even more important for a quintic spring in Figure \ref{OtherTuningAmplitudes2}(c).

\section{Conclusion}\label{conclusion}

Our purpose in this study is the development of a nonlinear absorber, the NLTVA, for mitigating the 
vibrations of a nonlinear resonance of a mechanical system. A specific objective is to ensure 
the effectiveness of the absorber in weakly as well as strongly nonlinear regimes of motion for which the 
primary system's resonance frequency can undergo substantial variations.

To this end, the additional design parameter offered by nonlinear devices, i.e., the mathematical form of the absorber's restoring force, is exploited
thereby synthesizing nonlinearity for enhanced performance. We show that, if the NLTVA is `a mirror' of the primary system, 
a nonlinear counterpart of Den Hartog's equal-peak method can be established. Simple, though accurate, analytic formulas
 are derived for this nonlinear equal-peak method. They lead to the design of an absorber with excellent performance 
 in a relatively large range of forcing amplitudes. Interestingly, the coupled system Duffing-NLTVA is found to exhibit dynamics that bear resemblance to 
 that of a linear system. 
 
For very strongly nonlinear regimes, inherently nonlinear dynamical instabilities, 
namely detached resonance curves and quasiperiodic solutions, appear. Despite these instabilities, the performance of the NLTVA remains always 
superior to the classical LTVA. Further research should study how these instabilities can be mitigated or eliminated. The extension of the present results to multi-degree-of-freedom 
primary structures will also be considered. Finally, another promising application of the NLTVA that 
deserves further investigation is the suppression of limit cycle oscillations \cite{Habib}, which exist, for instance, in aircraft wings and automotive disc brakes.

\section*{Acknowledgments}

The authors G. Habib, T. Detroux and G. Kerschen would like to acknowledge the financial support of the European Union (ERC Starting Grant NoVib
307265).

\end{document}